\documentclass[12pt]{article}
\usepackage{hyperref}
\usepackage{amsmath, amsfonts, amssymb,latexsym}
\usepackage{wasysym,graphicx,stmaryrd,tikz,bclogo}
\usepackage{pgfplots}

\usepackage{amsmath, amsfonts, amssymb,latexsym}

\input epsf.sty

\newtheorem{Theorem}{Theorem}[section]
\newtheorem{Lemma}{Lemma}[section]

\newcommand{\BEQ}{\begin{equation}}     % Gleichungen Anfang ..
\newcommand{\BEA}{\begin{eqnarray}}
\newcommand{\BD}{\begin{displaymath}}
\newcommand{\EEQ}{\end{equation}}       % .. und Ende
\newcommand{\EEA}{\end{eqnarray}}
\newcommand{\ED}{\end{displaymath}}
\newcommand{\del}{\delta}

\newcommand{\eps}{\varepsilon}          % epsilon

\newcommand{\R}{\mathbb{R}}

\newcommand{\N}{\mathbb{N}}

\renewcommand{\P}{\mathbb{P}}

\newcommand{\Id}{{\mathrm{Id}}}

\def\proba{{\mathbb{P}}}

\newcommand{\eop}{\hfill $\Box$}        % quod erat demonstrandum ...
\newcommand{\Medskip}{\medskip\noindent}
\newcommand{\Bigskip}{\bigskip\noindent}

\renewcommand{\Re}{{\rm Re\ }}          % Realteil
\newcommand{\half}{{1\over 2}}          % 1/2 als Bruch

\begin{document}

\title
{\bf Stoechiometric and dynamical autocatalysis for diluted chemical reaction networks}
\vskip -2cm
%\date\today
\date{}
\vskip -2cm
\maketitle

\vskip -2cm
\begin{center}
{\bf   J\'er\'emie Unterberger$^a$, Philippe Nghe$^b$}
\end{center}

%\maketitle

%\vskip 0.5 cm

\centerline {\small $^a$Institut Elie Cartan,\footnote{Laboratoire 
associ\'e au CNRS UMR 7502.} Universit\'e de Lorraine,} 
\centerline{ B.P. 239, 
F -- 54506 Vand{\oe}uvre-l\`es-Nancy Cedex, France}
\centerline{jeremie.unterberger@univ-lorraine.fr}

\centerline {\small $^b$UMR CNRS-ESPCI Chimie Biologie Innovation 8231,}
\centerline { ESPCI Paris, Universit\'e Paris Sciences Lettres,} 
\centerline{10 rue Vauquelin, 75005 Paris, France}
\centerline{philippe.nghe@espci.psl.eu}

\vspace{2mm}
\begin{quote}

\renewcommand{\baselinestretch}{1.0}
\footnotesize
{Autocatalysis underlies the ability of chemical and biochemical systems to replicate. 
Recently, Blokhuis et al.  (Blokhuis 2020) gave a {\em stoechiometric} definition of autocatalysis for reaction networks, stating the existence of a combination of reactions such that the balance for all autocatalytic species is strictly positive, and investigated minimal autocatalytic networks, called {\em autocatalytic cores}. By contrast,
  spontaneous autocatalysis -- namely, exponential
 amplification of all species internal to a reaction network, starting from a {\em diluted regime}, i.e. low
 concentrations -- is a {\em dynamical} property.
 
  We introduce
 here a topological condition (Top) for autocatalysis, namely: restricting the reaction network description to highly diluted species, we assume  existence of a strongly connected component possessing at least one reaction with multiple products (including multiple copies of a single species).
We find this condition to be necessary and sufficient for stoechiometric autocatalysis.
When degradation reactions have small enough rates, the topological condition further ensures {\em dynamical autocatalysis}, characterized by a strictly positive Lyapunov exponent giving   the  instantaneous exponential growth rate of the system. 

The proof is generally based on the study of auxiliary Markov chains.
We provide as examples general autocatalytic cores of Type I and Type III in the typology of  (Blokhuis 2020). 
In a companion article (Unterberger 2021), Lyapunov exponents and the  behavior in the growth regime 
are studied quantitatively beyond the present diluted regime .    
}
\end{quote}

\vspace{4mm}
\noindent

  
 \medskip
 \noindent {\bf MSC Classification (2020):} 34D08, 60J20, 80A32,
 92C42, 92C45, 92E20

\Medskip {\small P. Nghe acknowledges support from Institut Pierre-Gilles de Gennes (laboratoire d’excellence, “Investissements d’avenir” program ANR-10-IDEX-0001-02 PSL and ANR-10-LABX-31) and the European Research Council (ERC) under the European Union’s Horizon 2020 research and innovation programme (grant agreement No. [101002075]).}

\newpage
\tableofcontents

%\medskip

%%%%%%%%%%%%%%%%%%%%%%%%%%

\section{Introduction}

%%%%%%%%%%%%%%%%%%%%%%%%

\subsection{Our main result in a nutshell, and contexts for applications}

\noindent 
The chemical mechanism that epitomizes the ability of living systems to reproduce themselves is autocatalysis, namely, catalysis brought about by one of the products of the reactions. 
Autocatalysis must have been present from the early stages of the origin of life, 
from primitive forms of metabolism  (Allen 2019), to autocatalytic sets based on the first catalytic biopolymers (Kauffman 1986) and the emergence of sustained template-based replication of nucleic acids  (Eigen 1971).
Diverse artificial autocatalytic systems have been implemented in the laboratory (Hanopolskyi 2021), 
and remnants of ancestral autocatalytic networks may be found in extant metabolic networks (Kun 2008).
These examples reveal the diversity of autocatalytic mechanisms and chemistries.
However, the stoichiometry of autocatalytic has been characterized only recently  (Blokhuis 2020) and we still lack a systematic understanding of dynamical conditions for autocatalysis (Andersen 2020) which limits our ability to conceive plausible prebiotic scenarios (Jeancolas 2020). 

\Medskip To fill this gap, it is necessary to investigate how autocatalysis may emerge in complex mixtures.
This would help us understand the appearance of self-sustaining reactions in messy prebiotic mixtures (Danger 2020), and interpret experiments that search for such reactions (Berg 2019). 
Identifying autocatalytic systems is also critical to explain the appearance of Darwinian evolution, from complex mixtures (Danger 2020), to autocatalytic sets (Hordijk 2012) and ultimately template-based replication (Nghe 2015), a path which comprises multiple transitions and can be studied experimentally in RNA reaction networks (Ameta 2018, Ameta 2021).

\Medskip The focus here is on  {\em spontaneous} autocatalysis in chemical reaction networks, namely,
exponential amplification of a set of species with  low initial concentrations. This requires that certain other species, from which the network feeds, are provided in sufficiently large quantities in the environment. These resource species, sometimes called the 'food-set', may be constantly supplied from a large reservoir or external fluxes, or may be the products of reactions that already self-sustain in the milieu (Buss 1994). 

\Medskip Our main result, Theorem \ref{th:main}, gives a general condition, denoted (Top), for spontaneous autocatalysis to be possible in a {\em stoechiometric}, respectively  {\em dynamical} sense, understood as the existence of, respectively: combinations of reactions that lead to an increase of every autocatalytic species, and instantaneous growth of the dynamical system associated with the reaction network.  Our result holds provided that the reaction
set satisfies
the formal conditions stated in  (Blokhuis 2020): (i) {\em autonomy:} reactions should possess at least one reactant and one product; (ii) {\em non-ambiguity:} a species cannot be both a  reactant and  a product of the same reaction. 
Point (i) ensures that concentrations do not increase merely due to reactions that only consume species from the environment. Said differently, it ensures that any concentration increase depends
on the presence of another autocatalytic species, as required by the definition of autocatalysis  (Blokhuis 2020).
Point (ii) imposes a formal choice of coarse-graining in the description of the reaction network. This choice ensures that catalytic steps can be distinguished at the level of the stoechiometric matrix as the catalysts then appear in the stoichiometry (as shown in  (Blokhuis 2020)). Note that such a choice implies no restriction of generality, as it is always possible to introduce additional reaction intermediates in the description so that (ii) is respected  (Blokhuis 2020).

\begin{figure}
\label{fig:criterion}
\begin{center}
\begin{tikzpicture}[
scale=1.5,
roundnode/.style={circle, draw=black, thick, minimum size=10mm}, %fill=green!5
graynode/.style={circle, dashed, draw=black,fill=gray!20, thick, minimum size=10mm}, %fill=green!5
squarednode/.style={rectangle, draw=black, fill=gray!40, very thick, minimum size=10mm}
]
\node[roundnode] (a) at (0,0) {$a$};
\node[roundnode] (b) at (4,-2) {$b$};
\node[squarednode] (c) at (6,-2) {$c$};
\node[roundnode] (d) at (0,2) {$d$};
\node[roundnode] (e) at (5,1) {$e$};
\node[graynode] (f) at (3,2) {$f$};
\node[roundnode] (g) at (0,4) {$g$};
\node[graynode] (h) at (3,4) {$h$};
\node[roundnode] (i) at (4,4) {$i$};
\node[squarednode] (j) at (6,4) {$j$};
\draw[->, thick] (a)--(b);
\draw[<-, thick] (a) to[bend right=30] (1,1);
\draw[<-, thick] (d) to[bend left=30] (1,1);
\draw[-, thick] (1,1)--(e);
\draw[-, thick] (b) to[bend left=30] (5,-1);
\draw[-, thick] (c) to[bend right=30] (5,-1);
\draw[->, thick] (5,-1)--(e);
\draw[-, thick,dashed] (g) to[bend left=30] (1,3);
\draw[-, thick,dashed] (d) to[bend right=30] (1,3);
\draw[-, thick,dashed] (1,3)--(2,3) ;
\draw[->, thick,dashed] (2,3) to[bend left=30] (h);
\draw[->, thick,dashed] (2,3) to[bend right=30] (f);
\draw[->, thick] (d)--(g);
\draw[-, thick] (i) to[bend right=30] (5,3);
\draw[-, thick] (j) to[bend left=30] (5,3);
\draw[->, thick] (5,3)--(e);
\draw[->] (f)--(e);
\draw[->] (h)--(i);
\draw[->, thick] (g)  to[bend left=40] (i);
\end{tikzpicture}
\end{center}
\caption{Species $c$ and $j$ (gray squares) are initially abundant in the environment, thus can be safely ignored. The reaction $g+d \rightarrow h+f$ (dashed) has multiple reactants that are initially rare or absent, thus has a negligible rate compared to others and is discarded from the description. In the remaining graph, the set $\{a,b,d,e,g,i\}$ forms a strongly connected component (SCC), as there exists a directed path between any two of its members. Species $h$ and $f$ (dashed gray circles) are not part of the SCC. The SCC comprises a reaction ($e \rightarrow a+d$) with multiple products. Thus, the SCC is stoechiometrically autocatalytic (note that it is actually a Type III autocatalytic core according to  (Blokhuis 2020), see Supplementary Information \ref{SI:typeIII}). Furthermore, it is dynamically autocatalytic provided degradation rates of the species of the SCC are sufficiently small.}
\end{figure}
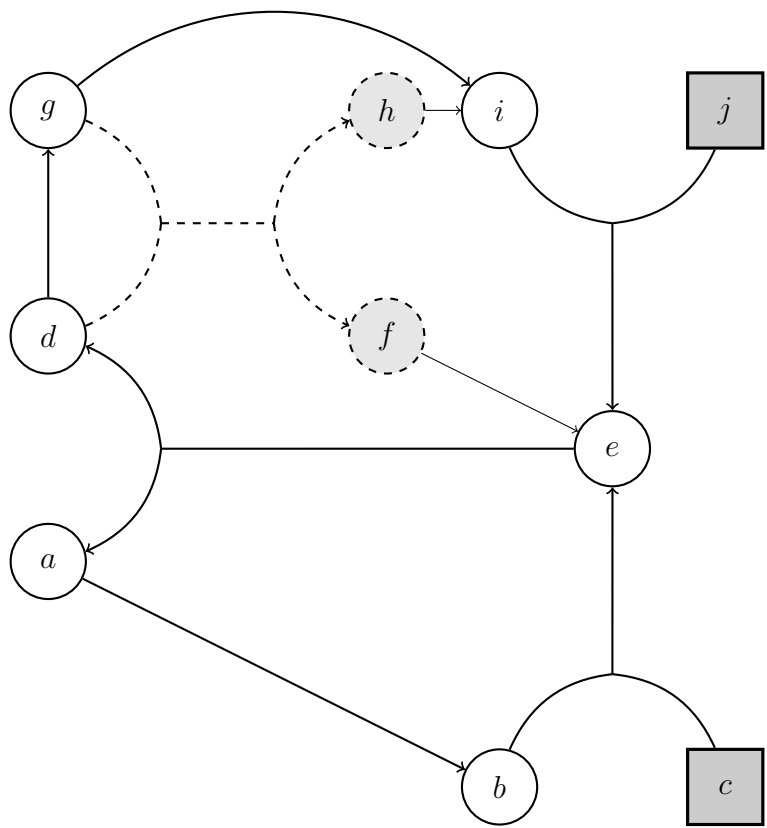

\Medskip Given the above conventions, verifying autocatalysis consists in isolating subsets of reactions that obey the topological criteria below (Fig. \ref{fig:criterion}): 
\begin{enumerate}
\item Retain only species that are initially absent or rare and discard from the description those that are abundant (the environment).
\item Dismiss reactions that have more than one reactant among the absent or rare species.
\item In the resulting network, identify strongly connected components which possess at least one reaction with multiple products within the component, including the case of multiple copies of a single species.
\end{enumerate}
Strongly connected components are defined as subgraphs in which any pair of vertices (species) are connected by a chain of reactions.
Successful verification of the steps above implies {\em stoechiometric autocatalysis}, independently of the reaction rates.
It further implies {\em dynamical autocatalysis} for sufficiently small degradation rates, as characterized by an exponential increase of every  species in the component assuming initially low concentrations, at least in the early phase of the dynamics.

%déplacé depuis 1.3
\vskip 1cm\noindent  Here is a brief outline of the article. In the remainder of this section 1, we introduce the formal framework for reaction networks. A simple motivating example is presented in section 2.  Sections 3 and 4 are the core of the article. We state our main result (Theorem \ref{th:main})
in section 3, and discuss equivalence of stoechiometric autocatalysis with the  topological property (Top). In section 4, we prove that this property implies  dynamical autocatalysis in the diluted regime when degradation reactions have small enough rates. We present perspectives for future work in section 5.  Finally,
section 6 provides supplementary information for the main text:
a presentation of type I and type III cycles, and mathematical
concepts and results used in the article, based on general Markov theory.

%%%%%%%%%%%%%%%%%%%%%%%%%

\subsection{Linearized dynamics of reaction networks}
\label{subsection:framework}

%%%%%%%%%%%%%

\Medskip We now introduce the framework of the present article, which is a mathematical and physical elaboration on the
recent theoretical work  (Blokhuis 2020)
by Blokhuis, Lacoste and Nghe on autocatalysis in chemical reaction networks, the main conclusions of which we first
recapitulate.

\Medskip The general setting is that of open reaction networks, see e.g.
(Esposito 2016) and references within. Chemical species fall into two categories:  {\em dynamical} (or {\em non-chemostatted}) species, whose concentrations vary over time according to kinetic (or stochastic, if present
in small number) equations, as opposed to {\em chemostatted} species, whose  concentrations are fixed (or large w.r. to dynamical species, so that their concentrations may be
considered as almost constant). 
Chemostatted species influence rates, but are not included
into the stoechiometric matrix (see below), therefore they need
not even be specified when dealing with stoechiometry alone.  In
 (Blokhuis 2020), only {\em autonomous} networks are considered,
i.e. every  reaction -- save {\em degradation reactions} --  is supposed to have at least one (dynamical)
reactant and at least one (dynamical) product.  Degradation
reactions $A\to \emptyset$ are natural in a biological
setting; they play a major r\^ole in the story, but are not taken explicitly into
account in the network.

%j'ai déplacé ce paragraphe
\Medskip The authors of  (Blokhuis 2020) further insist on the necessity of writing reactions in an {\em unambiguous} form,  i.e. in such a way
that  no 
chemical species can be both a reactant and a product of
a reaction. 
For instance, this avoids reactions to be written as $A+E\leftrightarrows B+E$ where the catalyst $E$ appears on both sides, thus cancels from the total stoechiometric balance.
Instead, the reaction should be written in two steps $A+E \leftrightarrows EA \leftrightarrows E+B$ which
formally ensures that $E$ appears in the stoechiometric balance and ultimately makes it possible to recognize the catalytic cycle associated with enzyme $E$ in the structure of the stoechiometric matrix (see  (Blokhuis 2020) for details). 

\Medskip The authors then 
introduce a {\em stoechiometric criterion for autocatalysis}
that depends only on the stoechiometric matrix ${\mathbb S}$,
a matrix with columns indexed by  reactions 
(other than degradation reactions)  $({\cal R}_1,\ldots,
{\cal R}_N)$, and rows indexed
by the set ${\cal S}=\{A_1,\ldots,A_{|{\cal S}|}\}$ of dynamical chemical species.
Recall that, by definition, each column of $\mathbb S$ corresponds to
the stoechiometry of a given reaction
\BEQ ({\cal R}): \qquad  s_{1} A_{i_1}+\ldots+s_{n} A_{i_n} \to s'_{1} A_{i'_1}+\ldots+s'_{n'} A_{i'_{n'}} 
\EEQ
that is, ${\mathbb S}_{j,{\cal R}}=-\sum_{\ell=1}^n s_{\ell}
 \del_{i_{\ell},j} + \sum_{\ell'=1}^{n'} s'_{\ell'} \del_{i'_{\ell'},j}$ (note that the coefficients of $\mathbb S$ depend on the choice of an orientation for every reaction).   
Namely, they require that there should exist 
\begin{itemize}
\item[(i)] a choice of orientations for reactions, and
\item[(ii)] a  positive reaction vector $c\in (\R_+)^{N}$ such
that ${\mathbb{S}}c>0$.
\end{itemize}
 This means that the reaction  obtained by taking the linear combination $\sum_{\cal R} c_{{\cal R}}
{\cal R}$ strictly increases the number of molecules of all species
in $\cal S$; in other terms, the chemical balance $({\mathbb S}c)_i$
for species $i$ is $>0$. The  main rationale for this condition
is Gordan's theorem (Adrian 2006), which  states that (ii) holds if and only
if there is no mass-like conservation law, i.e. there exists
no linear combination $\sum n_i [A_i]$ with positive coefficients
${\bf n}=(n_i)_{i\in {\cal S}}>0$ such that ${\bf n}\, \cdot\, {\mathbb S}=0$, i.e. preserved under all reactions.

\Medskip The authors go on to give a classification of all autocatalytic {\em cores}, that is of all {\em minimal
autonomous sub-networks} satisfying the above criteria, into
5 types I-V. (Types I and III are presented in Suppl. Info.)

\vskip 1cm \noindent Among foremost questions raised by this new classification, let us single out the two following:

\begin{itemize}
\item[(A)] {\em Are stoechiometrically autocatalytic networks
able to replicate ? Conversely, are chemical networks capable of
replication stoechiometrically autocatalytic ?}
\item[(B)] (If the answer to (A) is: yes, and assuming
some natural form for the rates, in particular, for
mass-action rates.) {\em Under which conditions over the concentrations and the rates does an autocatalytic network
indeed replicate ? If it does, can one estimate its replication rate ?}
\end{itemize}

\Bigskip  To be specific, {\em  kinetic rates will always be  assumed to be mass-action rates.}
 Then this work presents  an essentially complete answer to question (A) 
 in a specific regime which we call {\em diluted regime}, 
 where all concentrations of dynamical 
  species are {\em low}, and assuming that there are {\em no
  degradation reactions}, or that these have sufficiently
  small rates. The companion article (Unterberger 2021), on the other hand, presents a detailed case study for question (B) 
  for  a broad class of autocatalytic cores in a large part of
  the growth regime, well beyond the diluted regime, and
   in presence of degradation reactions; it rests
 on the notations and concepts introduced here, which are therefore
 presented in great generality. 
    
 \Medskip Partial answers to questions (A) and (B) are already
 available in  (Blokhuis 2020); they are based on self-consistent
 equations for survival probability, and are therefore rather given
 in the framework of stochastic networks, assuming only
 a few molecules are initially present. Generally speaking,  survival
 criteria are given in a form akin to that given by King (King 1982).   In (Unterberger 2021), it is proved that this regime has connections to the diluted regime in the kinetic framework studied
 here, in the case when the Lyapunov exponent is zero.  On the other hand, our formalism makes it possible to make predictions for
 the Lyapunov exponent far from the latter regime, as developed in the companion article (Unterberger 2021).
 
 \Medskip Despite the fact that kinetic equations are not linear in general, our work is
    largely based on the study of the  time-evolution of linear
     evolution models of the type
\BEQ  \frac{d[A_i]}{dt}=\sum_j M_{ij} [A_j]  \label{eq:M0} \EEQ
with negative diagonal coefficients $M_{ii} \le 0$ and 
positive off-diagonal coefficients $M_{ij} \ge 0, i\not =j$, which
are found in different contexts in the literature. 
%Je trouverais la phrase ci-dessous un peu confusante car on ne réfère pas ensuite à ce genre de modèle, je l'ai ré-écrite et ajouté la ref Eigen, mais je ne suis  même pas sûr que la phrase soit utile.
Note that these equations are formally similar to linear mutation-selection models, where
off-diagonal coefficients are interpreted as mutation rates, and
selection rates related to diagonal coefficients; see e.g. (Eigen 1971, Kussell 2005). 
Since we
are mainly inspired by Markov techniques, we speak here of $M$ as
{\em generalized adjoint Markov generator}, see Supp. Info. 
Indeed, when the sum of coefficients on any column is zero,
the total concentration $\sum_{i=1}^{|{\cal S}|} [A_i]$ is a constant.
Normalizing it to one, (\ref{eq:M0}) yields a master equation, namely, the time-evolution of a probability measure. On the 
other hand,
he case $|M_{j,j}|>\sum_{i\not=j} M_{ij}$ yields the
time-evolution of a sub-Markov process, i.e. a Markov process
with killing rates $a_j=|M_{j,j}|-\sum_{i\not=j} M_{ij}$. 
The case when some $a_j$ are negative -- indicating 'source'
terms -- is not standard in probability theory, but remains mathematically valid. 
Indeed, whatever the sign of  $a_j$, the {\em Feynman-Kac formula} yields the solution to (\ref{eq:M0}) in terms of a {\em sum over
paths with transition rates $w_{i\to j}$ proportional
to $M_{ji}$} (mind the index transposition due to the fact
that $M$ is a backward generator). Thinking in terms of
kinetic networks (and in spite of the fact that these are assumed to be  written as autonomous systems), a {\em positive killing rate} $a_j$ is associated
to a {\em degradation reaction} $A_j \to \emptyset$, whereas
a {\em negative killing rate} is associated to  an inverse
{\em creation reaction} $\emptyset \to A_j$. (Since  chemostatted species are left out
of the equations,  the  latter, seemingly
creation-ex-nihilo, reaction should be thought of really  as $A'\to A_j+A''$, where $A',A''$ are chemostatted species). 

\Bigskip Markov generators come out by {\em linearizing} the
kinetic equations. Formally, the time-evolution of
concentrations may be expressed in terms of the stoechiometric matrix and the current vector $J=(J_i)_{i=1,\ldots,n}$,
\BEQ \frac{d[A]}{dt}={\mathbb S}J.  \label{eq:3} \EEQ
Linearizing around given concentrations $([A_i])_{i=1,\ldots,|{\cal S}|}$, one gets for infinitesimal variations $[A]\longrightarrow [A]+A$ (mind the notation without square brackets for
{\em variations})
\BEQ \frac{dA}{dt}={\mathbb S}J_{lin}([A],A) \label{eq:4}
\EEQ
where $J_{lin,i}([A],A)=\sum_{\ell} J_{lin,i}^{\ell}([A]) 
A_{\ell}$ is linear in the variations. Letting
\BEQ M([A]):={\mathbb S}J_{lin}([A]),  \EEQ
we get the linear system
\BEQ \frac{dA}{dt}=M([A]) A. \label{eq:5} \EEQ

\Bigskip The matrix $M([A])$ 
 is sometimes (but not always)
a generalized Markov generator. 
 A case for which $M([A])$ is indeed a generalized Markov generator is when {\em each reaction has exactly
{\em one} reactant}, so that its rate is linear in its concentration:  the reaction $A_1\overset{k_+}{\longrightarrow} s_1 B_1+\ldots+ s_n B_n$, $n\ge 1$ makes the following additive contribution
to $M([A])$,

\begin{center}
\begin{tikzpicture}[scale=0.5]
\draw(1,1.5) node {$A_1$};
\draw(-1,0) node {$A_1$};
\draw(-1,-1) node {$B_1$};
\draw(-1,-2) node {$\vdots$};
\draw(-1,-3) node {$B_n$};
\draw(0.5,1)--(0,1)--(0,-7)--(0.5,-7);
\draw(1,0) node{$-k_+$};
\draw(1,-1) node{$ s_1 k_+$};
\draw(1,-2) node {$\vdots$};
\draw(1,-3) node {$s_n k_+$};
\draw(1,-4) node {$0$};
\draw(1,-5) node {$\vdots$};
\draw(1,-6) node {$0$};
\draw(4.5,-3) node {$0$};
\draw(8,1)--(8.5,1)--(8.5,-7)--(8,-7);
\end{tikzpicture}
\end{center}

An important particular case is that of a 1-1 reaction
$A_1   \overset{k_+}{\to} B_1$; the contribution to  $M([A])$ is
then simply \label{additive-contribution}

\begin{center}
\begin{tikzpicture}[scale=0.5]
\draw(1,1.5) node {$A_1$};
\draw(-1,0) node {$A_1$};
\draw(-1,-1) node {$B_1$};
\draw(0.5,1)--(0,1)--(0,-5)--(0.5,-5);
\draw(1,0) node{$-k_+$};
\draw(1,-1) node{$  k_+$};
\draw(1,-2) node {$0$};
\draw(1,-3) node {$\vdots$};
\draw(1,-4) node {$0$};
\draw(4.5,-2) node {$0$};
\draw(8,1)--(8.5,1)--(8.5,-5)--(8,-5);
\end{tikzpicture}
\end{center}

\noindent for which the sum of coefficients on the $A_1$-column is zero, in coherence with probability preservation.

\Medskip   Interestingly, autocatalytic cores, as proved in  (Blokhuis 2020), satisfy the latter condition --  
 except that the stoechiometry is more general, allowing for reactions of type $s A  {\longrightarrow} s_1 B_1+\ldots+ s_n B_n$, $s\ge 1$. This only turns the top coefficient $M_{A,A}$ into $-sk_+$. The associated killing rate for species $A$ is $(1-(s_1+\ldots+s_n))k_+$, it is $\le 0$ for reactions of the type  $A\overset{k_+}{\longrightarrow} s_1 B_1+\ldots+ s_n B_n$, $n\ge 1$ (but not
 necessarily when $s\not=1$). 
 
\Medskip
Going one step further,  we note that reactions with $\ge 2$
reactants have a vanishing rate in the limit when concentrations go to zero. In that limit, furthermore, 
all killing rates are $\le 0$.  We call this the {\em zero
concentration
limit} of networks. In this limit, where the linearized time evolution generator involves only mutations and creation reactions, it is easily conceived
that autocatalysis should hold in any reasonable sense.  {\em The
adjacency graph associated to the generalized Markov generator $M([A]=0)$  (see  section \ref{section:3} for its precise
construction)
is denoted $G({\mathbb S})$;  it depends only on the stoechiometric matrix. }

\vskip 1cm

%{\centerline{**********************************************}}
\subsection{Diluted regime and statement of condition (Top)}
\label{subsection:TOP}
%PN: Je propose de faire une sous-section ici. Sinon on n'arrive pas vraiment à se repérer vu que ça se répète un peu. Avec une sous-section, le problème n'apparaîtra pas trop.
%\vskip 1cm

\Bigskip The present study is devoted  to {\bf diluted
 networks}. These are systems with {\em low},
but nonzero, concentrations, for which reactions with $\ge 2$ reactants exist but have low rate compared to the others. 
The physical picture is that of a system of reactions of three types:
\begin{itemize}
\item[(i)] reversible reactions, with linear rates, involving {\em one} reactant
and {\em one} product,
\BEQ A_i \leftrightarrows A_j;  \label{eq:6} \EEQ
\item[(ii)]  irreversible forward reactions 
involving {\em one} reactant
and {\em several} products, with linear rates,
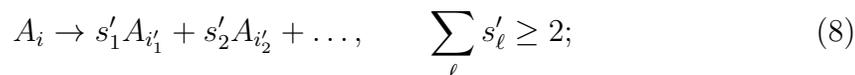
\BEQ A_i\to s'_1 A_{i'_1}+ s'_2 A_{i'_2}+\ldots, \qquad 
\sum_{\ell} s'_{\ell}\ge 2;  \label{eq:7}
\EEQ
such reactions are totally irreversible in the zero concentration limit;
\item[(iii)] and, possibly, the reverse reactions associated to the reactions in (ii),
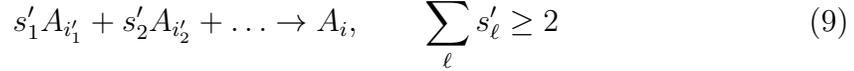
\BEQ s'_1 A_{i'_1}+ s'_2 A_{i'_2}+\ldots \to A_i,\qquad 
\sum_{\ell} s'_{\ell}\ge 2 \label{eq:8}
\EEQ
with nonlinear, but low (compared to (i) and (ii)) or zero reaction rates. 
\end{itemize}
Degradation reactions (which are non-autonomous) may also
be included. {\bf Degradationless diluted networks} are
diluted networks in which degradation reactions are either absent or
have small enough rates.

\Medskip In that setting, it is natural to approximate
reactions (iii) by their linearizations, which have in any case a small rate compared to reactions of type (i) or (ii). We do not get
in general a generalized adjoint Markov generator, because
off-diagonal coefficients of $M([A])$ are not necessarily positive. (They {\em are} positive when reverse reactions are
strictly of the form $mA_{i'}\to A_i$ with $m\ge 2$; 
see  Suppl. Info. for examples and general statements). Even in that case, however, the Feynman-Kac formula holds (see Suppl. Info.), so our general strategy works.

\Medskip Generally speaking, reactions such as (\ref{eq:6}) or (\ref{eq:7})
may be represented in the form  of a {\em hypergraph} called
{\em hypergraph associated to ${\mathbb S}$}  
 (Anderson 2019, \S 2), with 'pitchforks' connecting $A_i$ to $A_{i'_{\ell}}$  by $s'_{\ell}$ arrows in the case of a
one-to-several irreversible reaction. Under these conditions, a natural quantity characterizing the replication rate is the {\bf Lyapunov exponent} $\lambda_{max}\equiv \lambda_{max}(M([A]))$,  by definition 
\BEQ \lambda_{max}:=\max\{ \Re (\lambda)\ |\ \lambda\  {\mathrm{eigenvalue\
of\ }}  M([A])\}. \EEQ
Under our hypotheses,  it can be proved (using 
Perron-Frobenius theorem) that $\lambda_{max}$ is 
an eigenvalue of $M([A])$ with multiplicity $1$, and that 
an eigenvector associated to $\lambda_{max}$ can be chosen 
in such a way that all its coordinates are $>0$. 
  When $\lambda_{max}$ is positive, it characterizes the onset of the exponential growth regime of the system, namely, for small initial concentrations,
\BEQ \frac{ [A_j](t)}{\max_i \, ([A_i](t=0))}  \approx  e^{\lambda_{max}t} \EEQ
for all species $A_j\in{\cal S}$, for  not-too-large time values $t$.  When $\lambda_{max}>0$, we say that
the reaction network is {\bf (strongly) autocatalytic in the dynamical sense}  -- or
{\bf (strongly) dynamically autocatalytic} --.  Generalizing (in order to include the case of reducible
networks, see below), we say that the reaction network is {\bf weakly autocatalytic in the
dynamical sense} (or {\bf weakly dynamically autocatalytic}) if $\lambda_{max}>0$ is an eigenvalue
of $M([A])$ (multiplicity can be arbitrary, and Jordan blocks associated to $\lambda_{max}$ may be
non-trivial), and an eigenvector $v$ associated to $\lambda_{max}$ may be chosen in such a way that
all its coordinates are $\ge 0$. Then (11) remains valid for all species $A_j$ such that $v_j > 0$.

\Medskip Turning now to the stoechiometric side, in coherence with the above discussion,
we neglect altogether reverse reactions (iii): we
say that {\em a diluted network is stoechiometrically autocatalytic if there exists a positive reaction vector $c$ such that ${\mathbb S}c>0$, where the orientation of reversible, type (i) reactions is arbitrary, but   forward reactions (ii) are  given  positive orientation.}

\Medskip We can now state  our main result (see Theorem \ref{th:main} below); recall that $G({\mathbb S})$ is the
adjadency graph associated to the generalized Markov
matrix $M([A]=0)$. Thinking of it as if it were the graph
of a conventional Markov matrix, we decompose it into
 classes (see Suppl. Info. \S \ref{subsection:GM}), with
 probability flow flowing downstream  from minimal
 to maximal classes.  Letting ${\cal C}$ be one of the  classes, we now define its set of {\bf internal reactions}. If ${\cal R} \, :\, A\to s_1 A'_1+\cdots + s_n A'_n$ is an irreversible reaction such that $A\in{\cal C}$, and 
$\{i=1,\ldots,n \ |\ A'_i\in {\cal C}\}=\{1,\ldots,n'\}$ for
some $n'\ge 1$, then we introduce the {\bf truncated reaction} ${\cal R}_{{\cal C}}\, :\, A\to s_1 A'_1+\cdots + s_{n'} A'_{n'}$. Internal reactions of $\cal C$ are then: (i) reversible
reactions $A\leftrightarrows A'$ with $A,A'\in{\cal C}$; (ii)
truncated reactions ${\cal R}_{{\cal C}}\, :\, A\to s_1 A'_1+\cdots + s_{n'} A'_{n'}$ as above, with $A,A'_1,\ldots,A'_{n'}\in {\cal C}$. Note that, if $n'=1$, we obtain a reaction
of a new type: an {\em irreversible 1-1 reaction}.  
 All other irreversible reactions, i.e.
reactions of the form ${\cal R}\, :\, A\to s_1 A'_1+\cdots + s_n A'_n$ with $A\in {\cal C}$ and $A'_1,\ldots,A'_n\not\in 
{\cal C}$  are {\bf purely external reactions}; such reactions
can appear only in reducible networks.

\Medskip What we prove is the following: 
\begin{enumerate}
\item  {\em diluted networks are stoechiometrically autocatalytic
if and only if the following topological condition on the hypergraph associated to ${\mathbb S}$ holds,}

\Medskip (Top)\, : \qquad  {\bf all minimal classes of $G({\mathbb S})$ contain
at least one internal one-to-several irreversible reaction},

\Medskip  i.e. each minimal class $\cal C$ contains a truncated 
reaction  ${\cal R}_{{\cal C}}\, :\,  A\to s_1 A'_1+\cdots + s_{n'} A'_{n'}$ with $s:=\sum_{i=1}^{n'} s_i\ge 2$;
\item {\em the latter topological condition implies  weak dynamical autocatalysis in the diluted regime, i.e. for small enough
concentrations, if there are no degradation reactions 
 nor purely external reactions, or,
 more generally, if the rates of those are small enough. Strong dynamical autocatalysis
holds 
in the specific case of  an irreducible network   in absence of degradation reactions, or if the rates of
those are small enough.}  
\end{enumerate} 

\Medskip When $G({\mathbb S})$ is {\em irreducible}, condition 1. 
states that at least one irreversible forward reaction (\ref{eq:7})
must be present in the system for stoechiometric autocatalysis
to hold -- an obvious statement in view of Gordan's theorem since
${\bf n} \, \cdot\, {\mathbb S}=0$ if ${\bf n}=\left(\begin{array}{c} 1 \\ \vdots\\ 1 \end{array}\right)$ and only $1-1$  reversible reactions
(\ref{eq:6})  are present in the system --, and most importantly,
that this is a sufficient condition. The presence of  irreversible forward reactions is also necessary for dynamical autocatalysis to hold -- otherwise only mutation-like coefficients are present, kinetic equations are
those of a conventional Markov system, and then it is known that
all generator eigenvalues have $\le 0$ real part --, but
2. states again that this is a sufficient condition.

\Medskip When $G({\mathbb S})$ is {\em reducible}, it is easy to see that the
condition (Top) on minimal classes in 1. is necessary. Namely, let $\cal C$ be one of the minimal classes, and ${\mathbb S}_{\cal C}$ the stoechiometric matrix associated with  species
in $\cal C$ and   internal reactions of $\cal C$. If
all internal reactions are 1-1, then the same argument based
on Gordan's theorem implies that, for every positive reaction vector $c_{{\cal C}}$, the balance $({\mathbb S}_{{\cal C}} \,
c_{\cal C})_i$ is $\le 0$ for at least one of the species $i\in{\cal C}$. 
Now, including {\em external} (for $\cal C$) reactions ${\cal R}\, :\, A_i\to s_1 A'_1+\ldots+ s_n A'_n$ with
$A_i\in {\cal C}$ and  $A'_1,\ldots,A'_n\not\in{\cal C}$ can only worsen the balance for species $A_i$, while other reactions $A'\to  s_1 A'_1+\ldots+s_n A'_n$ with $A'\not\in{\cal C}$ -- and
therefore, $A'_1,\ldots,A'_n\not\in{\cal C}$ also since $\cal C$ is minimal -- do not change it. Thus, for example, the network
with hypergraph

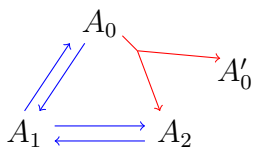
\begin{center}
\begin{tikzpicture}[scale=2]
\draw(0,0.732) node {$A_0$};
\draw(-0.5,0) node {$A_1$}; \draw(0.5,0) node {$A_2$}; 
\draw[->,blue] (-0.1,0.6)--(-0.4,0.15);
\draw[<-,blue] (-0.2,0.6)--(-0.5,0.15);

\draw[red] (0.15,0.65)--(0.25,0.55);
  \draw[->,red] (0.25,0.55)--(0.4,0.15); \draw[->,blue](-0.3,0.05)--(0.3,0.05);  \draw[<-,blue](-0.3,-0.05)--(0.3,-0.05); 

\draw[->,red] (0.25,0.55)--(0.8,0.5);
\draw(0.9,0.4) node {$A'_0$};
\end{tikzpicture}
\end{center}

\noindent featuring two reversible reactions $A_0\leftrightarrows A_1, A_1\leftrightarrows A_2$ (in blue) and 
an irreversible reaction ${\cal R}\, :\, A_0\to A_2+A'_0$ (in red)  coupling the minimal class ${\cal C}=\{A_0,A_1,A_2\}$ to
another class ${\cal C}'=\{A'_0\}$,  does not satisfy (Top),
because  the truncated irreversible reaction ${\cal R}_{\cal C} \, : \, A_0 \to A_2$ is 1-1.

\Medskip Looking closely at the (generalized) eigenspace associated to $\lambda_{max}$ in the reducible case,
one realizes that a complete discussion of the nature of dynamical autocatalysis (e.g. multiplicity
of $\lambda_{max}$, and support of an associated eigenvector $v$, i.e. the set of species $A_j$
such that $v_j > 0$) cannot rely only on topological considerations; see detailed example
p. \pageref{example:red}. In addition, the presence of reverse reactions (\ref{eq:8}), even at negligible rates from a
biological or chemical point of view, modifies the above characteristics.
Let us explain this subtlety on two minimal examples. Assume there are two classes ${\cal C},{\cal C}'$ with probabilistic
flow flowing out of ${\cal C}$ into ${\cal C}'$, 
\begin{center}
\begin{tikzpicture}[scale=0.5]
\draw(0,0) node {$\cal C$}; 
\draw[->](0,-0.5)--(0,-1.5);
\draw(0,-2) node {${\cal C}'$};
\end{tikzpicture}
\end{center}
and that, as a first case, (i) {\em only the maximal (downstream) class ${\cal C}'$ contains an internal irreversible reaction} (so that (Top) is not satisfied); then (excluding reverse reactions
going upstream from ${\cal C}'$ to ${\cal C}$) the network restricted to ${\cal C}'$ is irreducible and (strongly)
dynamically autocatalytic; thus the whole network is weakly dynamically autocatalytic.
In this case however, whatever ${\cal C}$ reactants present in the solution disappear exponentially
in time in favor of species in ${\cal C}'$. Now imagine choosing one of the reactions (8) connecting
${\cal C}$ to ${\cal C}'$ and adding the reverse reaction with a negligible rate $O(\eps)$, $0 <\eps\ll 1$. This makes
the network irreducible, implying strong dynamical autocatalysis, while perturbing only
slightly $\lambda_{max}$. The associated positive eigenvector $v = v(\eps)$ -- unique up to normalization
-- will have nonzero but very small coefficients along $\cal C$, making it probably difficult in
practice to observe exponential increase of the corresponding species. Next, consider as
a second case the possibility that (ii) {\em (Top) is satisfied, so that the network restricted to
$\cal C$} (i.e. suppressing all ${\cal C}'$-products of reactions with reactant in $\cal C$) {\em is autocatalytic, but ${\cal C}'$
contains no irreversible reaction, hence is not autocatalytic}. Then the network (as proved
in section \ref{section:4}) is already strongly autocatalytic in itself.

\Medskip Dismissing these subtleties and summarizing, our result may be rephrased as follows:  {\em stoechiometric
autocatalysis implies dynamical autocatalysis} in 
our diluted regime  and, in 
absence of degradation reactions; and  (at least in the case of irreducible networks), it may be said that the converse
is also true. 

\Bigskip {\em Remark.} As follows from the above, (Top) depends 
on the hypergraph associated to $\mathbb S$, not only on the graph $G({\mathbb S})$. However, the main tools in the proof are
based on properties of $G({\mathbb S})$.

%%%%%%%%%%%%%%%%%%%%%%%%%%%%%%%%%%%%%%%ù

\section{A motivating example: the simplest autocatalytic core}
\label{section:simplest}

%%%%%%%%%%%%%%%%%%%%%%%%%%%%%%%ùù

We treat in this section the simplest type I autocatalytic core
in the classification of  (Blokhuis 2020). It involves
two chemostatted species $(A,A')$, which may be thought of as
a redox or energy carrier (ATP/ADP) couple, or as fuel and waste
(Esposito 2019); two dynamical species $(B,B_1)$; and
   two reactions

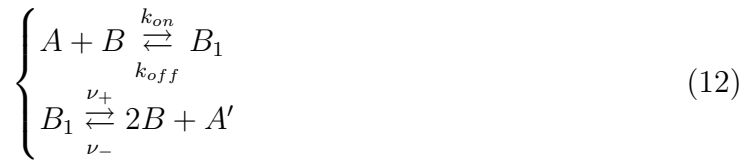
\BEQ \begin{cases} A + B \overset{k_{on}}{\underset{k_{off}}{\rightleftarrows}} B_1 \\
B_1 \overset{\nu_+}{\underset{\nu_-}{\rightleftarrows}}
2B+ A' \end{cases}
\EEQ

\Medskip
  Autocatalysis is made possible by the duplication reaction $B_1 \overset{\nu_+}{\to} 2B+A'$. 
 We also include degradation reactions
 
 \BEQ \begin{cases} B \overset{a_0}{\to} \emptyset \\
 B_1 \overset{a_1}{\to}\emptyset 
   \end{cases}
  \EEQ

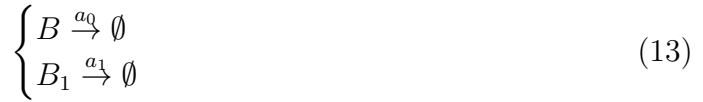
\Medskip  The degradationless diluted regime which is
the main topic of the
article is defined by  
\begin{itemize}
\item[(i)] (low concentrations) $[B],[B_1]\ll 1$. 
Kinetic equations lack any reference concentration or volume
to produce adimensional quantities, and chemostatted quantities $[A],[A']$ are
not limited,  so (by simple rescaling of the concentrations)
this criterion is equivalent to
\BEQ \nu_-[B]\ll 1.  \label{eq:2:low-concentrations} \EEQ
In other words, the reverse of the duplication reaction
is rate-limited.

\item[(ii)] (no degradation) $a_0,a_1=0$. Our
analysis actually extends (by perturbation) to low enough
degradation rates, 
\BEQ a_0,a_1\ll 1.   \label{eq:2:no-degradation} \EEQ

\end{itemize}

\Medskip{\em Kinetic equations are:}

\BEA  (\frac{d}{dt}+a_0) [B]=2\nu_+ [B_1] - (k_{on} [A][B]
-k_{off} [B_1]) - 2\nu_- [B]^2 [A'] \\
(\frac{d}{dt}+a_1) [B_1]=-\nu_+ [B_1] + (k_{on} [A][B]
-k_{off} [B_1])  + \nu_- [B]^2 [A']  
\EEA

When $\nu_-=0$, these equations are linear, otherwise
we linearize around $([B],[B_1])$, and find the  system
\BEQ \frac{d}{dt} \left(\begin{array}{c} B \\ B_1
\end{array}\right)=M\left(\begin{array}{c} B \\ B_1
\end{array}\right)
\EEQ
where $(B,B_1)$
 is an  infinitesimal variation  around
$([B],[B_1])$
, and
\BEQ M=\left[ \begin{array}{cc} -k_{on} [A] - 4\nu_- [A'] [B] - a_0 & 
k_{off} + 2\nu_+ \\ k_{on} [A] + 2\nu_- [A'] [B] & -k_{off}
-\nu_+ - a_1\end{array}\right] \label{eq:M22}
\EEQ
 Note that off-diagonal elements of $M$ are $> 0$, so
that, by the Perron-Frobenius theorem, the spectrum of $M$ consists of 
 two complex numbers $\lambda_{max}(M),\lambda_{min}(M)$
 with $\lambda_{max}(M)$ real, and $\lambda_{max}(M)>\Re\lambda_{min}(M)$. Furthermore, $M$ has an eigenvector for
 the eigenvalue $\lambda_{max}(M)$ 
 with positive coefficients.  Write $M=\left[\begin{array}{cc} -a & b \\ c & -d \end{array}
\right]$. Explicit computations actually produce
 two real numbers,
 \BEQ \lambda_{max}(M)=\half\Big(-(a+d)+\sqrt{(a+d)^2-4\det(M)} \Big)
 = \half\Big(-(a+d)+ \sqrt{(a-d)^2+ 4bc}\Big) 
 \EEQ
 and
 $\lambda_{min}(M)= \half\Big(-(a+d)- \sqrt{(a-d)^2+ 4bc}\Big)$.

\begin{Lemma} (see (England 2019) )
Let $M=\left[\begin{array}{cc} -a & b \\ c & -d \end{array}
\right]$ $(a,b,c,d>0)$ and $\lambda_{max}=\lambda_{max}(M)$ the
eigenvalue of $M$ with largest real part. Then the following alternative holds, 
\begin{itemize}
\item[(i)] If  $\det(M)= ad-bc<0$, then
 $\lambda_{max}>0$;
\item[(ii)] if $\det(M)=0$, then $\lambda_{max}=0$;
\item[(iii)] if $\det(M)>0$, then $\lambda_{max}<0$.
\end{itemize}
\end{Lemma}

\Medskip  Autocatalysis is then equivalent to  the
condition $\det(M)<0$.  We now check that, in the
degradationless diluted regime defined by (\ref{eq:2:low-concentrations},
\ref{eq:2:no-degradation}), 
\BEQ \det(M)=-k_{on}[A]\nu_+ + O(\nu_-[B])+O(a_0)+O(a_1)
<0.
\EEQ

\Medskip   Going beyond this particular regime,  autocatalysis 
is not the rule. Let us consider two specific cases:

\begin{enumerate}
\item[(i)] (no reverse reaction) We neglect  reverse reactions
by setting $k_{off}=0$ and $\nu_-=0$. Then
\BEQ \det(M)=-2k_{on}[A] \nu_+ +  (k_{on}[A]+ a_0)( \nu_+ +  a_1)
<0
\EEQ
if and only if (see King's criterion
 (King 1982)) the product of the specificities
 of positively oriented reactions along the replication cycle
 $B\to B_1\to 2B$ is
 larger than $\half$,
 
\BEQ \frac{k_{on}[A]}{k_{on}[A]+a_0}  \frac{\nu_+}{\nu_++a_1}>\half.
\EEQ

\item[(ii)] (no degradation)  We assume here that  $a_0=a_1=0$. Then
\BEQ \det(M)=-k_{on} [A] \nu_+  +  2k_{off} \nu_- [A'][B] <0  \EEQ
 if and only if 

\BEQ  [B]<  [B]_{max}:= \frac{k_{on}[A]}{k_{off}} \, \times\, \frac{\nu_+}{2\nu_-[A']},\EEQ
or equivalently,
\BEQ \frac{k_{on}[A]}{k_{off}} \, \times\, \frac{\nu_+}{2\nu_-[A'][B]}>1,
\EEQ
a criterion somewhat analogous to King's criterion, but featuring 
the ratio (product of forward reaction rates)/(product of reverse
reaction rates). 

\end{enumerate}

%%%%%%%%%%%%%%%%%%%%%%%%%%%

%%%%%%%%%%%%%%%%%%%%%%%%%

\section{(Top) characterizes stoechiometric  autocatalysis in diluted networks}
\label{section:3}

%%%%%%%%%%%%%%%%%%%%%%%%%

We reconsider in this section the stoechiometric autocatalysis criterion of Blokhuis-Lacoste-Nghe  (Blokhuis 2020) in
the case when all concentrations are {\em low}.
Under such circumstances, reactions involving $>1$ reactants
have very small rate. It is therefore reasonable to  discard them from the beginning when dealing with the stoechiometric
definition of autocatalysis. 

\Medskip This leads us to introduce a subclass
of stoechiometrically autocatalytic networks, which we call
{\bf diluted stoechiometrically  autocatalytic networks}. Consider a reaction network with
species set $\cal S$ and reaction set $\{1,\ldots,N\}$,  
choose an orientation for each reaction, characterizing
{\em forward} reactions, by opposition to {\em reverse}
reactions. Degradation reactions possibly exist, but are not 
included in the reaction set, and play no r\^ole in the
discussion.  Choosing some arbitrary ordering of reactions,
we get a stoechiometric matrix ${\mathbb S}$.  Then we require the following conditions:

\begin{itemize}
\item[(i)] the reaction network is unambiguous and autonomous;
\item[(ii)] there exists a  positive reaction vector $c\in (\R_+)^{N}$ such
that ${\mathbb{S}}c>0$;
\item[(iii)] {\bf each reaction has only one reactant, and its stoechiometry is $1$.}
\end{itemize}

\noindent The last condition (iii) restricts the class introduced in
 (Blokhuis 2020). Recall however (as already mentioned) that all autocatalytic cores
satisfy partially this condition, in the sense that
all reactions have exactly one reactant (with arbitrary stoechiometry, however). 

\Medskip {\bf Diluted networks, associated graph.}  Removing
assumption (ii), we get the definition of a {\bf diluted network}, which is the general class of topological networks
of interest in the present work. We
 associate to such a reaction network its {\bf split graph}
 (or simply {\bf graph}) $G({\mathbb S})$, which
depends only on the stoechiometric matrix, and corresponds mathematically to
the linearization of the kinetic network in the zero concentration limit. It may be defined topologically
as follows: (direct) reactions of the type $A\to s_1 B_1 +
\ldots + s_n B_n$ $(n\ge 1, s_1,\ldots,s_n\in\N^*)$ such that
$s_1+\ldots+s_n\ge 2$, i.e. with $>1$ products, are totally irreversible in the limit of vanishing concentrations, therefore they contribute to $G({\mathbb S})$ irreversible
arrows 
\BEQ A\to B_1,\ldots, A\to B_n \EEQ
 upon {\bf splitting} the reaction
into reactions with unique products. On the other hand, forward reactions
of the type $A\to B$ with only one product are reversible; therefore,
they contribute to $G({\mathbb S})$ reversible arrows $A\leftrightarrows B$. In case of multiple  arrows $A\to B$, we only keep one, in
order not to have multiple edges from $A$ to $B$. 
This happens if there are several competing irreversible reactions $A\to
s B+ s_2 B_2+\ldots + s_n B_n$, $A\to s'B+ s'_2 B'_2+\ldots+s'_{n'} B'_{n'}$,
or if
  irreversible reactions $A\to sB+\cdots$ and a
reversible reaction $A\leftrightarrows B$ coexist.  We always assume that $G({\mathbb S})$ is {\em connected} (otherwise one can reduce the analysis to each
of the subsystems defined by the connected components).

\Medskip Having a graph instead of a hypergraph with pitchforks
connecting several reactants and several products (see below, and examples in
\S 5.1 and 5.2) is a major simplification. To be precise, we
note that $G({\mathbb S})$ is sometimes not quite enough to
caracterize  stoechiometric autocatalysis: in case an irreversible reaction
$A\to sB+\cdots$ and the reversible reaction $A\leftrightarrows B$ coexist (so that $A$  and $B$ are in the 
same class ${\cal C}$, see below), the graph $G({\mathbb S})$ by itself does not keep track of the
existence of the irreversible reaction. Then we keep the memory  of the 
irreversible transition $A\to B$ by saying that ${\cal C}$ contains an {\em internal irreversible reaction}. In case $A\to B+\cdots$
is not in competition with a reversible reaction, but $A$ and $B$
are in the same class thanks to the presence of an irreversible
reaction $B\to A+\cdots$, both split reactions $A\to B$ and $B\to A$ are
considered as internal irreversible reactions. A simple way to
summarize these rules is to decide that {\em reversible arrows are
painted blue, irreversible arrows are painted red, and 
red prevails.} Thus we get a graph with {\em two-colored edges}. This is sometimes useful, but still not enough to define our
topological condition (Top) when the graph is not
irreducible (see \S \ref{subsection:TOP}).  Classes are defined below without taking
the color of the arrows into account.

\Medskip {\bf Classes.}  Upon linearizing the
time-evolution equations,  while neglecting reverse reactions, one obtains  a {\em generalized Markov matrix} (see Suppl. Info.) $M$
with graph $G({\mathbb S})$. This justifies resorting to the
usual description of $G$ in terms of communicating classes, connected by irreversible arrows.  Arrows
define a partial order of classes, with ${\cal C'}>{\cal C}$ if  there is a path from 
${\cal C}$ to ${\cal  C}'$, i.e. if ${\cal C}'$
is downstream of ${\cal C}$. 
In Suppl. Info. (\S \ref{subsection:GM}), the reader will find several examples
worked out in details: cores of type I and III,

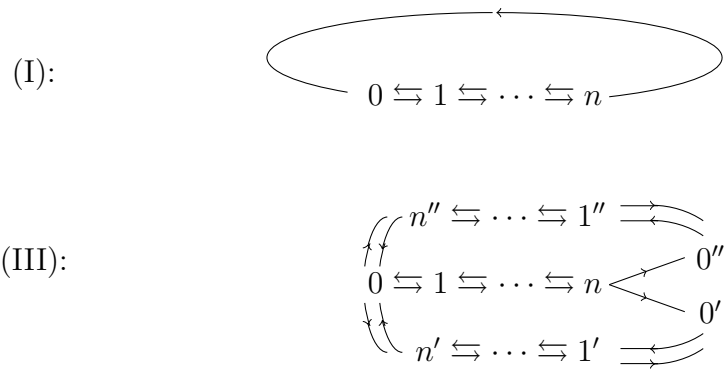
\Bigskip
{\centerline{
\begin{tikzpicture}
\draw(-6,0.3) node {(I):};
\draw(0,0) node{$0 \leftrightarrows 1 \leftrightarrows \cdots
 \leftrightarrows n$};
\draw[->](1.65,0) arc(-60:90:3 and 0.6);
\draw(0.1,1.12) arc(90:230:3 and 0.6); 
\begin{scope}[shift={(0,-2.5)}]
\draw(-6,0.3) node {(III):};
\draw(0,0) node{$0 \leftrightarrows 1 \leftrightarrows \cdots
 \leftrightarrows n$};
\draw[->](1.65,0)--(1.95+0.2,0.18); \draw(1.95+0.2,0.18)--(2.255+0.4,0.36);
\draw[->](1.65,0)--(1.95+0.2,-0.18); \draw(1.95+0.2,-0.18)--(2.255+0.4,-0.36);
%\draw[->](2.25,0)--(2.25+0.4*0.6,0.3*0.6);
%\draw(2.25+0.4*0.6,0.3*0.6)--(2.25+0.8*0.6,0.6*0.6);
%\draw[->](2.25,0)--(2.25+0.4*0.6,-0.3*0.6);
%\draw(2.25+0.4*0.6,-0.3*0.6)--(2.25+0.8*0.6,-0.6*0.6);
\draw(3,0.36) node {$0''$}; \draw(3,-0.36) node {$0'$};
\draw[->](2.9,0.65) arc(45:90:1 and 0.7);
\draw(2.3,0.85)--(1.8,0.85);
\draw[-<](2.9,0.2+0.65) arc(45:90:1 and 0.7);
\draw(2.3,0.2+0.85)--(1.8,0.2+0.85);
\draw(0.3,0.9) node{$n'' \leftrightarrows   \cdots
  \leftrightarrows 1''$};
\draw[->](-1.1,0.9) arc(90:140:0.35 and 1.2); 
\draw(-1.33,0.55) arc(140:165:0.35 and 0.8); 
\draw[-<](-0.1-1.2,0.9) arc(90:140:0.35 and 1.2); 
\draw(-0.1-1.43,0.55) arc(140:165:0.35 and 0.8); 
%
%\draw[->](2.9,-0.65) arc(-45:-90:1.2 and 0.6);
\draw[->](2.9,-0.65) arc(-45:-90:1 and 0.7);
\draw(2.3,-0.85)--(1.8,-0.85);
\draw[-<](2.9,-0.2-0.65) arc(-45:-90:1 and 0.7);
\draw(2.3,-0.2-0.85)--(1.8,-0.2-0.85);
\draw(0.3,-0.9) node{$n' \leftrightarrows   \cdots
  \leftrightarrows 1'$};
\draw[->](-1.1,-0.9) arc(-90:-140:0.35 and 1.2); 
\draw(-1.33,-0.55) arc(-140:-165:0.35 and 0.8); 
\draw[-<](-0.1-1.2,-0.9) arc(-90:-140:0.35 and 1.2); 
\draw(-0.1-1.43,-0.55) arc(-140:-165:0.35 and 0.8); 
\end{scope}
\end{tikzpicture}
}}

\Bigskip
 and the
"$A_1 A_2 A_3 \longrightarrow B_1 B_2 B_3$" autocatalytic kinetic reaction network, and its graph $G_{(123)\to (1'2'3')}$, where $(A_1,A_2,A_3)$, resp. $(B_1,B_2,B_3)$ are
encoded by indices $(1,2,3)$, resp. $(1',2',3')$: 

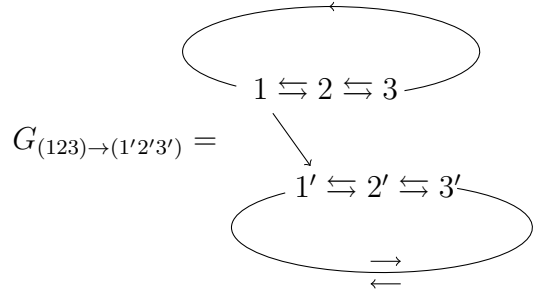
\bigskip
{\centerline{
\begin{tikzpicture}
\draw(-2.8,-0.7) node {$G_{(123)\to (1'2'3')}=$};
\draw(0,0) node{$1 \leftrightarrows 2 \leftrightarrows 3$};
\draw[->](1.05,0) arc(-60:90:2 and 0.6);
\draw(0.1,1.12) arc(90:230:2 and 0.6); 
\draw[->](-0.7,-0.3)--(-0.2,-1);
\draw(0.7,-1.3) node{$1' \leftrightarrows 2' \leftrightarrows 3'$}; 
\draw(1.05+0.7,0-1.3) arc(60:-230:2 and 0.6);
\draw[->](0.7+0.1-0.2,-1.3-1.12+0.15)--(0.7+0.1+0.2,-1.3-1.12+0.15);
\draw[<-](0.7+0.1-0.2,-1.3-1.12-0.15)--(0.7+0.1+0.2,-1.3-1.12-0.15);
\end{tikzpicture}
}}

\Bigskip Note that the stoechiometry is not indicated, nor
is it important in the analysis that follows, once understood
that irreversible arrows come from splitting reactions
with $>1$ products. As a matter of fact,  Type (I) cores ($B_0,\ldots,B_n)$ have
originally a "pitchfork" reaction $B_n\to 2B_0$

\medskip
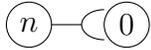
\begin{tikzpicture}
\draw(0,0) circle(0.3);
\draw(0,0) node {$n$};
\draw(0.3,0)--(0.7,0);
\draw(0.7,0) arc(180:90:0.3 and 0.2); 
\draw(0.7,-0) arc(-180:-90:0.3 and 0.2);
\draw(1.3,0) circle(0.3);
\draw(1.3,0) node {$0$};
\end{tikzpicture} 

Type (III) cores, involving species $A_i$, $i=0,\ldots,n$,
$B'_{i'}$, $i'=0',\ldots,n'$, $B''_{i''}$, $i''=0'',\ldots,n''$, have originally a pitchfork

\medskip
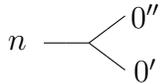
\begin{tikzpicture}
\draw(1.3,0) node {$n$};
\draw(1.65,0)--(1.95,0); \draw(1.95,0)--(2.25,0);
\draw(2.25,0)--(2.25+0.4*0.6,0.3*0.6);
\draw(2.25+0.4*0.6,0.3*0.6)--(2.25+0.8*0.6,0.6*0.6);
\draw(2.25,0)--(2.25+0.4*0.6,-0.3*0.6);
\draw(2.25+0.4*0.6,-0.3*0.6)--(2.25+0.8*0.6,-0.6*0.6);
\draw(3,0.36) node {$0''$}; \draw(3,-0.36) node {$0'$};
\end{tikzpicture}

\Bigskip Only one-sided arrows indicate the location
of the original hypergraph pitchforks. The one-sided arrow $n\to 0$
in (I)  indicates any reaction $B_n\to m B_0$ with $m=2,3,\ldots$. The one-sided arrows $A_n\to B''_{0''}, A_n\to B'_{0'}$ come either from $A_n\to s'' B''_{0''}+s'B'_{0'}$, $s',s''=1,2,\ldots$ or from $(A_n\to m'' B''_{0''}, A_n\to
m' B'_{0'})$, $m',m''=2,3,\ldots$, or from a combination of these. 

\Bigskip
   All cores are irreducible. The
"$A_1 A_2 A_3 \longrightarrow B_1 B_2 B_3$"  network, on the other
hand, 
has two classes, ${\cal C}=(1,2,3)$ and ${\cal C}'=(1',2',3')$, with ${\cal C}'$ downstream of ${\cal C}$. 
The partial ordering defines in particular
minimal (upstream) and maximal (downstream) classes;
here, ${\cal C}$ is minimal, and ${\cal C}'$ is maximal.   

\Bigskip  Our main result is the following:

\begin{Theorem} \label{th:main}
\begin{enumerate}
\item A diluted network  with stoechiometric matrix $\mathbb S$ is a diluted stoechiometrically autocatalytic
network if and only if the following topological condition {\em (Top)} of the hypergraph associated to $\mathbb S$ is satisfied:

\Medskip {\em  (Top)\, : \qquad   each of the minimal classes of $G({\mathbb S})$  contains  at least
one internal one-to-several  irreversible reaction.}

\item A diluted network  satisfying {\em (Top)} is weakly autocatalytic in the dynamical
sense if there are no degradation reactions, or, more generally, if their rates are small enough. Furthermore, it is strongly
autocatalytic in the dynamical sense under the same
conditions if the network is irreducible.
\end{enumerate}
\end{Theorem}

\Bigskip We prove in the rest of the section the first part of the Theorem, concerning {\em stoechiometric autocatalysis}; the second part will be proved in the next section. 

\Medskip Stoechiometric autocatalysis, at least in the case
of an irreducible network, can  be proven quite simply by playing directly with the columns of  the stoechiometric matrix ${\mathbb S}$; see  Suppl. Info. \ref{SI:Perron}. 
Instead, we provide here a general demonstration using properties of $G({\mathbb S})$. Though a little more
involved, it has the advantage of exploiting the properties
of an underlying {\em auxiliary} Markov chain, which will
also play a major r\^ole in \S \ref{section:4}. In the case of a reducible network, arguments rely on the class decomposition of the  graph
 $G({\mathbb S})$.

\Medskip We have already proved in the Introduction that (Top) is necessary for a diluted
network to be autocatalytic. 
So the interesting part is to show that (Top) is a {\em sufficient condition} for
autocatalysis.  We split
the proof into several points. The general idea is to construct an explicit
reaction vector $c$  which depends on the choice
of a kinetic rate for each reaction, and is a perturbation of
the stationary flow vector for an auxiliary Markov chain. 

\Medskip  The chemical balance for species
$A_k$ associated to a reaction ${\cal R}\ :\: s_1 A_{i_1}+
\ldots+ s_n A_{i_n}\to s'_1 A_{i'_1}+\ldots+s'_{n'} A_{i'_{n'}}$
will be denoted $\del_{\cal R}[A_k]= -\sum_j s_j\del_{k,i_j} + \sum_{j'} s'_{j'} 
\del_{k,i'_{j'}}$. Then the total chemical balance  for species
$A_k$ associated
to the combination of reactions represented by the reaction vector $c$   is $\del[A_k]=\sum_{\cal R} c_{\cal R}\,  \del_{\cal R}[A_k]$.

%%%%%%%%%%%%ù

\Medskip  {\bf A. (stationary flows for split graph).} Theorem 
\ref{th:main} (1) is obtained by perturbation from
the following remark. One can define an  {\bf auxiliary Markov chain}
  $(\tilde{X}(t))_{t\ge 0}$ (a conventional, continuous-time Markov chain, i.e. with vanishing killing rates) with transition rates 
 $\tilde{k}_{i\to j}$ obtained by superposing the
 following transitions:
 
\begin{itemize}
\item[(i)] 
Reversible transitions with rates $k_{i\to j}, k_{j\to i}$ are associated to 1-1 reactions
 of the type ${\cal R}_{i\to j}:\ \  A_i \overset{k_{i\to j}}{\to} A_j, \ {\cal R}_{j\to i}:\ \  A_j \overset{k_{j\to i}}{\to} A_i$;
 
 \item[(ii)] 
Irreversible transitions with rates $s_j k_i^+$are associated to split irreversible 1-1 reactions $\tilde{\cal R}:
 \ \ A_i \overset{s_j k^+_{i}}{\to} A_j$, $j=i_1,\ldots,i_n$  coming from the one-to-several
 irreversible reaction $A_i\overset{k_i^+}{\to} s_{i_1} A_{i_1}+\ldots+s_{i_n} A_{i_n}$.
 
\end{itemize}

\noindent The associated adjoint Markov generator is obtained
by summing matrices with only two non-vanishing coefficients
as on p.\pageref{additive-contribution}; then the sum of coefficients on any column is zero, which ensures probability
preservation.
In other words, $\tilde{k}_{i\to j}=\sum_{{\cal R}\, :\, A_i\to A_j}
\tilde{k}_{i\to j}(\tilde{\cal R})$, where, depending on the split
reaction $\tilde{\cal R}\, :\, A_i\to A_j$, one has defined:  $\tilde{k}_{i\to j}(\tilde{\cal R})=k_{i\to j}$ (one-to-one reaction) or $s_j k^+_i$ (split forward reaction $A_i\to A_j$ coming
from a one-to-several reaction $A_i\to s_j A_j+\cdots$) or $0$ (excluded
reverse reaction).

\Medskip Assume the graph $G({\mathbb S})$ is irreducible.
Then the auxiliary Markov chain $(\tilde{X}(t))_{t\ge 0}$ is irreducible;  it reproduces
correctly the transition rates of the kinetic network from $A_i$ to $A_{i_{\ell}}$ for irreversible transitions (ii), but
{\em increases} the exit rate from $A_i$, since $\frac{d[A_i]}{dt}=
-sk_i^+ [A_i]$ (by probability conservation) with $s=\sum_{\ell} s_{i_{\ell}}\ge 2$  for the Markov chain, as compared to $\frac{d[A_i]}{dt}=-k_i^+[A_i]$ for the
kinetic network.  { The auxiliary Markov chain admits
exactly one stationary probability measure $\mu=(\mu_i)_{i=1,\ldots,|{\cal S}|}$.  Define $\tilde{c}_{i\to j}(\tilde{\cal R}):=
\begin{cases} \mu_i \tilde{k}_{i\to j}(\tilde{\cal R}) \qquad {\mathrm{if}}\ \tilde{\cal R}\, :\, i\to j \\ 0 \qquad {\mathrm{else}}
\end{cases}$, and   let $\tilde{c}_{i\to j}=\sum_{\tilde{\cal R}\, :\, i\to j} \tilde{c}_{i\to j}(\tilde{\cal R})= \mu_i \tilde{k}_{i\to j}  $ be the stationary flow along the edges. 
  Then the antisymmetrized quantity $\tilde{J}_{i\to j}:=
  \sum_{\tilde{\cal R}} \tilde{J}_{i\to j}(\tilde{\cal R})\equiv 
  \sum_{\tilde{\cal R}} \Big\{\tilde{c}_{i\to j}(\tilde{\cal R})-
  \tilde{c}_{j\to i}(\tilde{\cal R})\Big\}=\tilde{c}_{i\to j} - 
  \tilde{c}_{j\to i}$ is the associated current, and {\em the
total current for species $i$ vanishes by stationarity, i.e.
hence 
\BEQ \sum_j \tilde{J}_{i\to j}=0. \label{eq:stationarity}
\EEQ  }

\Medskip {\bf Choice of the reaction vector c.}  Going
back to the initial network, {\em we now define $c({\cal R}):=\tilde{c}_{i\to j}({\cal R})=\mu_i k_{i\to j}$  for the reversible one-to-one reaction ${\cal R}\, :\, A_i\overset{k_{i\to j}}{\to} A_j$, and $c({\cal R}):=\mu_i k_i^+$  for the irreversible one-to-several
reaction ${\cal R}\, :\, A_i \overset{k_i^+}{\to} s_{i_1}A_{i_1}+\ldots+s_{i_n}A_{i_n}$.}  Note that (for convenience) we
have chosen to accept both orientations for reversible 1-1
reactions; this is equivalent to choosing an orientation for
each of them and letting $c({\cal R}_{(i,j)}):=c({\cal R}_{i\to j})-c({\cal R}_{j\to i})$ (the orientation may be chosen in such a way that $c({\cal R}_{(i,j)})\ge 0$).  The chemical balance $\del [A_i]:=({\mathbb S}c)_i$ for species $i$  is obtained by summing
\BEQ -\tilde{c}_{i\to j}({\cal R})+\tilde{c}_{j\to i}({\cal R})=-\tilde{J}_{i\to j}({\cal R}) \EEQ
for a reversible one-to-one reaction ${\cal R}$ connecting
$i$ and $j$, 
\BEQ -c({\cal R})=-\mu_i k_i^+ \EEQ
for the reactant of a one-to-several reaction ${\cal R}\, :\, A_i\to \cdots$, and
\BEQ +s_{i}\mu_j k_j^+ = +\tilde{c}_{j\to i}(\tilde{\cal R}) \EEQ
for products of a one-to-several reaction ${\cal R}\,:\, A_j\to s_i A_i+\cdots$, split into several 1-1 reactions including $\tilde{\cal R}\, :\, A_j\to s_i A_i$. By construction, we obtain
\BEQ \del [A_i]=-\sum_j \tilde{J}_{i\to j}
\EEQ
if  species $i$ is {\em not} the reactant
of a one-to-several reaction; thus, in that case,
$\del [A_i]=0$. If, on the other hand, $i$ is the reactant
of a  one-to-several reaction $A_i  \overset{k_i^+}{\to} s_{i_1}A_{i_1}+\ldots+s_{i_n}A_{i_n}$ with associated split 
reactions $\tilde{\cal R}_{i,i_{\ell}}\, :\, A_i \to A_{i_{\ell}}$, then the associated balance for $[A_i]$ 
is 
\BEQ -\mu_i k_i^+>-\sum_{\ell} \tilde{c}_{i\to i_{\ell}}(\tilde{\cal R}_{i,i_{\ell}})=
-\mu_i sk_i^+ \EEQ
 with $s=\sum_{\ell} s_{i_{\ell}}>1. $  Comparing with 
 the above stationarity equation (\ref{eq:stationarity}), we may
 conclude:   
{\em our choice for the vector $c$ yields a strictly
positive balance for reactants
of a one-to-several reaction, and zero balance for all
other species.} 

\Medskip{\em Remark.} If the graph is over-connected, i.e. if
reversible 1-1,  or one-to-several  irreversible, reactions can be removed  without breaking irreducibility, then
the auxiliary Markov chain may be defined  while leaving them out,
yielding  another simpler set of coefficients $c_{\cal R}$ that vanish for left-out reactions.

%%%%%%%%%%%%%%%%%%%%

\Bigskip  {\bf B. (irreducible networks).} The reaction vector $c$ constructed in {\bf A.} is not quite satisfactory yet. We now turn to a {\em perturbation argument
for irreducible networks}, ensuring that there exist  vectors
$\del c^q=\left(\begin{array}{c} (\del c^q)_{{\cal R}_1}  \\
\vdots\\  (\del c^q)_{{\cal R}_N} \end{array}\right)$, $q=1,2,\ldots$
vanishing for $q$ large enough such that 
${\mathbb S} (c+\sum_{q\ge 1}\eps^q \del c^q)>0$ for all small enough
$\eps>0$. By hypothesis, there exists at least one irreversible
reaction. Choose one, ${\cal R}_0:A_0 \overset{k_0^+}{\to} s_{i_1}A_{i_1}+\ldots+s_{i_n}A_{i_n}$, and define $c\equiv c(\mu)$ as in the previous paragraph, $\mu$ being the stationary probability measure for $\tilde{X}$. Since the 
balance for $[A_0]$ is $>0$, we can tilt $\mu$ by a small amount in direction $0$, i.e. replace $\mu_0$ by $\mu_0+\eps$,
while keeping $\del[A_0]>0$. This is equivalent to saying
that $c \rightsquigarrow  c+\eps \del c^1$, with
$\del c^1({\cal R}_0)=k_0^+$, yielding
 $(c+\eps\del c^1)({\cal R}_0)=(\mu_0+\eps)k_0^+$; similarly,
 $\del c^1({\cal R})=k_{0\to j}$, resp. $(k'_0)^+$ for all other
 possible reactions $A_0\overset{k_{0\to j}}{\to} A_j$ or 
 $A_0 \overset{(k'_0)^+}{\to} s'_{i'_1}A_{i'_1}+\ldots+
 s'_{i'_{n'}} A'_{i'_{n'}}$ with reactant $A_0$;   and 
 $\del c^1({\cal R}')=0$ for all other
reactions. But then $\del[A_{i_{\ell}}]$ is
shifted by $+ \eps s_{i_{\ell}} k_0^+$, and possibly
other positive coefficients $(+\eps k_{0\to i_{\ell}}$ or $+\eps s'_{i_{\ell}} (k'_0)^+)$, so the balance for 
species $0$ and 
for all products of ${\cal R}_0$ is now $>0$; more precisely,
$\del [A_0]$ is of order $\eps^0$, while $\del [A_{i_{\ell}}]$, $\ell=1,\ldots,n$ -- and similarly, the balance for
all products of reactions with reactant $A_0$ --  are of order $\eps^1$.

\Medskip  We now let ${\cal S}_0:=\{0\}$, define
 ${\cal S}_1\subset{\cal S}$ to be made up of $0$, together
 with all products of reactions having $0$ as reactant, and  consider
products of reactions  having as reactant one of the elements
of the set ${\cal S}_1\setminus {\cal S}_0$. Since
the graph is irreducible, the corresponding set of reactions
can be empty only if ${\cal S}_1={\cal S}$. If this is not the case, tilt $\mu$ by a small uniform amount in all directions
indexed by the set ${\cal S}_1\setminus{\cal S}_0$, i.e. replace $\mu_i$ by
$\mu_i+\eps^2$ for all $i\in {\cal S}_1\setminus{\cal S}_0$.  For convenience, we reindex the set of species so that ${\cal S}_1\setminus{\cal S}_0=\{1,\ldots\}$. Choosing one
of the above reactions, either one-to-several ${\cal R}_1: A_1 \overset{k_1^+}{\to}
s_{i_1} A_{i_1}+\ldots+s_{i_n}A_{i_n}$ or one-to-one,
${\cal R}_{1\to j}:A_i \overset{k_{i\to j}}{\to} A_j$, this
is equivalent to saying that $c\rightsquigarrow 
c+\eps\del c^1+\eps^2 \del c^2$, with $\del c^2({\cal R}_1)=k_1^+$, resp. $\del c^2({\cal R}_{1\to j})=k_{1\to j}$. We thus
shift $\del [A_i]$, $i\in {\cal S}_1\setminus{\cal S}_0$, by $-O(\eps^2)$, 
and simultaneously $\del [A_{i'}]$ ($i'$ ranging in the set of products
of reactions having as reactant one of the elements of ${\cal S}_1\setminus{\cal S}_0$, including possibly species in ${\cal S}_1$) by $+O(\eps^2)$. The $\del [A_i]$  were of order $\eps^0$, resp. 
$\eps^1$ at previous step for $i\in {\cal S}^0$, resp.
${\cal S}^1\setminus{\cal S}^0$; the $\eps^2$-corrections do not change these
orders, but ensure that now $\del [A_i]$, $i\in {\cal S}_2\setminus{\cal S}_1$ are of order $\eps^2$, where ${\cal S}_2\setminus{\cal S}_1$ is the set of new products.    
We stop the induction in $q$  as soon as we have 
exhausted all species, i.e. the maximum index $q$ is the
minimum index such that ${\cal S}_q={\cal S}$. 

\Medskip {\em A simple example.} Consider the network with
species $A_0,A_1,A_2$, reversible 1-1 reactions $A_0\leftrightarrows A_2$
and $A_1\leftrightarrows A_2$, and a single irreversible one-to-several reaction ${\cal R}_0\, :\, A_0\to A_1+A_2$ with $s=2$. The
graph is

 
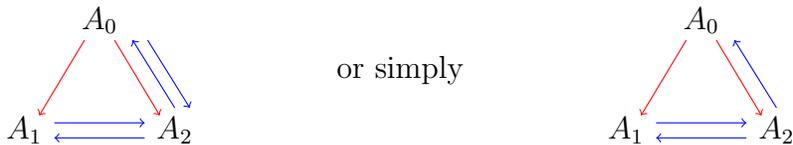
\begin{tikzpicture}[scale=2]  \draw(0,0.732) node {$A_0$};
\draw(-0.5,0) node {$A_1$}; \draw(0.5,0) node {$A_2$}; 
\draw[->,red] (-0.1,0.6)--(-0.4,0.1);  \draw[->,red] (0.1,0.6)--(0.4,0.1); \draw[->,blue](-0.3,0.05)--(0.3,0.05);  \draw[<-,blue](-0.3,-0.05)--(0.3,-0.05); 
\draw[->,blue](0.5,0.15)--(0.22,0.6); 
\draw[<-,blue](0.6,0.15)--(0.32,0.6);
\draw(2,0.4) node {or simply};
\begin{scope}[shift={(4,0)}]
\draw(0,0.732) node {$A_0$};
\draw(-0.5,0) node {$A_1$}; \draw(0.5,0) node {$A_2$}; 
\draw[->,red] (-0.1,0.6)--(-0.4,0.1);  \draw[->,red] (0.1,0.6)--(0.4,0.1); \draw[->,blue](-0.3,0.05)--(0.3,0.05);  \draw[<-,blue](-0.3,-0.05)--(0.3,-0.05); 
\draw[->,blue](0.5,0.15)--(0.22,0.6); 
%\draw[<-,blue](0.6,0.15)--(0.32,0.6);
\end{scope}
\end{tikzpicture}

\smallskip \noindent following the convention that "red prevails". The network
is irreducible. Choose all rates to be equal to 1. Then
the adjoint Markov generator  of the auxiliary  chain is
\BEQ \left(\begin{array}{lcc} -2 & & \\ 1  & & \\ 1 & &
\end{array}\right) + \left(\begin{array}{ccc} - 1 & & 1\\
 \\ 1 & & -1 \end{array}\right)+ \left(\begin{array}{ccc}
  \\ & -1 & 1 \\ & 1 & -1 \end{array}\right)=\left(\begin{array}{ccc} -3 & 0 & 1 \\ 1 & -1 & 1 \\ 2 & 1 & -2 \end{array}\right).
  \EEQ
Stationary measures are multiples of $\mu:=\left(\begin{array}{c} 1 \\ 4 \\ 3 \end{array}\right)$.  Stationary flows are
$\tilde{c}_{0\to 1}=1,\tilde{c}_{0\to 2}=2; \ \tilde{c}_{1\to 0}=0,\tilde{c}_{1\to 2}=4; \ \tilde{c}_{2\to 0}=3,\tilde{c}_{2\to 1}=3$, and then stationary currents are $\tilde{J}_{0\to 1}=1, \tilde{J}_{0\to 2}=-1, \tilde{J}_{1\to 2}=1$.  Following
our construction, we choose for reaction vector $c$ with
$c({\cal R}_0)=1$ and $c({\cal R}_{0\to 2})=1,c({\cal R}_{2\to 0})=3, c({\cal R}_{1\to 2})=4,c({\cal R}_{2\to 1})=3$. Then
\BEQ \del[A_1]=c({\cal R}_0)-c({\cal R}_{1\to 2})+c({\cal R}_{2\to 1})=0 ;
\EEQ
similarly, $\del[A_2]=0$; and $\del[A_0]=-c({\cal R}_0)-c({\cal R}_{0\to 2}) + c({\cal R}_{2\to 0})=1$, which can be identified
with $(s-1)\mu_0 k_0^+$ using the notations of the proof. 
We perturb it by a one-step construction since ${\cal S}_1=\{0,1,2\}$: we replace $c$ by $c+\eps \del c^1$ with 
$\del c^1({\cal R}_0)=\del c^1({\cal R}_{0\to 1})=\del c^1({\cal R}_{0\to 2})$.  Thus  the perturbed balance $\del [A_0]=1-\eps, \del [A_1]=+\eps, 
\del [A_2]=+2\eps$ is $>0$ for all species as soon as $0<\eps<1$. 

\Bigskip {\bf C. (reducible networks).} We must finally
adapt the above argument to the case of a reducible network.
To have a picture in mind, the reader may think of the
"contracted graph" 

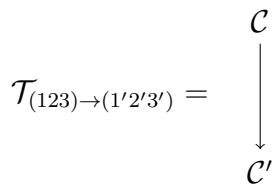
{\centerline{\begin{tikzpicture}
\draw(-2,-1) node {${\cal T}_{(123)\to (1'2'3')}=$};
\draw(0,0) node {$\cal C$};
\draw(0,-2) node {${\cal C}'$};
\draw[->](0,-0.3)--(0,-1.7);
\end{tikzpicture}}}

of the "$A_1 A_2 A_3 \longrightarrow B_1 B_2 B_3$"   network (see
\S \ref{subsection:GM}),
or, for a more general example, 
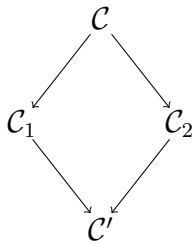
\Medskip

{\centerline{\begin{tikzpicture}[scale=0.7]
\draw(0,0) node {$\cal C$};
\draw[->](-0.2,-0.3)--(-1.3,-1.7); \draw(-1.5,-2) node {${\cal C}_1$};
\draw[->](0.2,-0.3)--(1.3,-1.7); \draw(1.5,-2) node {${\cal C}_2$};
\draw(0,-4) node {${\cal C}'$};
\draw[->](-1.3,-2.3)--(-0.2,-3.7);
\draw[->](1.3,-2.3)--(0.2,-3.7);
\end{tikzpicture}}}

In both examples here, there is a unique minimal class, ${\cal C}$, and a unique maximal class, ${\cal C}'$. Note that 
arrows go downwards, defining a probability flow from
minimal classes to maximal classes. We define
the {\em height} $h({\cal C}'')$ of a class ${\cal C}''$ to be the minimal distance on the contracted graph from  a 
minimal class to it. Here e.g.  $h({\cal C})=0,h({\cal C}')=1$ 
on our first example, and $h({\cal C})=0, h({\cal C}_1)=h({\cal C}_2)=1,h({\cal C}')=2$ on our second example. Our proof is by induction on the maximal height $h_{max}$. The
case $h_{max}=0$ has been solved in {\bf B.}, so we assume
$h_{max}\ge 1$.

\Medskip The argument goes as follows. Consider
a minimal class ${\cal C}$ connected downwards to ${\cal C}_1,
\ldots,{\cal C}_m$. A reaction ${\cal R}$ is {\em internal} to ${\cal C}$ if its reactant and all its products belong to ${\cal C}$; one then writes ${\cal R}: {\cal C}\to {\cal C}$. On the other hand, irreversible arrows from ${\cal C}$ to ${\cal C}_i$ represent split irreversible reactions $\tilde{\cal R}\,:\, A\to A_i$
with $A\in {\cal C}$ and $A_i\in {\cal C}_i$, coming from the
linearization of a one-to-several reaction ${\cal R}\, :\, A\to s_1 A'_1+\ldots +
s_n A'_n$. There are two cases:
\begin{itemize}
\item[(i)] (purely external reaction) either  $A'_i,i=1,\ldots,n$ all belong to $\uplus_{j=1}^m
{\cal C}_j$, so that all split reactions $\tilde{\cal R}\, :\, A
\to A'_i$ are external;
\item[(ii)] (mixed reaction) or one of the $A'_i$ belongs to ${\cal C}$, so that
$\tilde{\cal R}\, :\, A\to A'_i$ is an internal irreversible
reaction of ${\cal C}$.
\end{itemize}
The second case is called a mixed case because some of the $A'_i$ belong
to ${\cal C}$, and some do not, hence the one-to-several reaction ${\cal R}$ is neither internal nor external. Now, if there is no mixed
reaction with reactant in ${\cal C}$, we can extract from the
set of reactions those which are internal to ${\cal C}$, and
build the ${\cal C}$-valued auxiliary Markov chain $(\tilde{X}_{{\cal C}}(t))_{t\ge 0}$ as in {\bf A.}  with set of transitions associated to  those internal reactions.  The construction in {\bf A.} and {\bf B.}
 yields a positive
vector $c_{{\cal C}}=(c_{\cal R})_{{\cal R}\, :\, {\cal C} \to 
{\cal C}}$ such that the associated chemical balance
for all species in ${\cal C}$ is $>0$. 

\Medskip Considering now the case of a mixed reaction ${\cal R} \, :\, A\to s_1 A'_1+\ldots+s_n A'_n$ with $A'_1,\ldots,A'_{n'}\in {\cal C}, (A'_{\ell})_{\ell>n'}\in \uplus_{j=1}^m {\cal C}_j$, we split it for our purposes into a truncated internal reaction ${\cal R}_{{\cal C}} \, :\, A\to s_1 A'_1+\ldots+s_{n'} A'_{n'}$, and 
$n-n'$ external split reactions $\tilde{\cal R}\, :\, A\to A'_i$, 
$i=n'+1,\ldots,n$. Joining truncated internal reactions ${\cal R}_{{\cal C}}$ to the set of internal reactions, one proceeds
as in the previous paragraph, and obtains a positive vector
$c_{{\cal C}}=(c({\cal R}))_{{\cal R}\, :\, {\cal C}\to {\cal C}}$,
where now $ {\cal R}\, :\, {\cal C}\to {\cal C}$ represents
the set of all (truncated or not) reactions internal to $\cal C$,
such that the associated chemical balance
for all species in ${\cal C}$ is $>0$.

\Medskip We proceed similarly
for all minimal classes.

\Medskip  Consider  now a height 1 class ${\cal C}_1$.  Start as in
the previous paragraph by constructing a ${\cal C}_1$-valued
auxiliary Markov chain with set of transitions associated to  the  (truncated or not) reactions internal to ${\cal C}_1$. Proceed similarly for all
classes of height $1$. Using the construction in {\bf A.} and coupling with the height 0 class
reaction vectors obtained in the previous step, one obtains 
a reaction vector $c=(c_0,c_1)$ such that $c_0({\cal R})>0$, resp.
  $c_1({\cal R}) >0$  iff
${\cal R}$ is (truncated or not) internal to a height 0, resp. 1 class, and the associated balance is $>0$, resp. $\ge 0$, for species belonging
to height 0, resp. height 1 classes. 

\Medskip We now adapt the perturbation argument of {\bf B.} 
First,  if ${\cal R}\, :\, A\to \cdots$, $A$ belonging to
a minimal class $\cal C$, is of mixed type, we redefine $c(
{\cal R})=c_{{\cal C}}({\cal R}_{{\cal C}})$. Choosing a class
 ${\cal C}'$ of height 1,  we now
explain how to obtain a strictly positive balance for species  in 
${\cal C}'$. There are two cases:

\begin{itemize}
\item[(i)] (purely external case) Assume that all reactions 
${\cal R}\, :\, A\to s' A'+\cdots$, such that  $A'\in {\cal C}'$ and 
$A$ in a class of  height 0, are purely external, so none of 
these have been taken into account previously in the auxiliary
Markov chains.  The balance associated to such  reactions is strictly negative for the  reactant $A$, 
and strictly positive for  products, including $A'$. Choosing
a small enough coefficient
$c({\cal R})$ for them,  the net balance
for height 0 species remains $>0$, and we get a strictly
positive balance  for $A'$. 
\item[(ii)] (mixed case) Assume there exists a mixed reaction 
${\cal R}\, :\, A\to (s_1 A'_1+\ldots+s_{n'} A'_{n'}) + A'+ \cdots$, with
$A,A'_1,\ldots,A'_{n'}$ in a height 0 class ${\cal C}$, and  $A'\in {\cal C}'$. This reaction  has already been taken into account,
by construction $c_{{\cal C}}({\cal R}_{{\cal C}})>0$. Replacing
the truncated internal reaction ${\cal R}_{{\cal C}}$ by
${\cal R}$ only increases the balance for external species,
including $A'$. 

\end{itemize}

\Medskip In both cases, one has obtained a positive balance for
at least one species in each
height 1 class, which can be considered as a local influx. 
 One may now modify
the construction in  {\bf B.} by simply using
the local influx (instead of the positive balance due to an internal irreversible reaction) to perturb $c_1$, and obtains a positive vector $c'$ such that
$c'({\cal R})=c({\cal R})$ if the reactant of ${\cal R}$ belongs to a height 0 class, and the balance associated to $c'$ is $>0$ for species belonging to classes of height $\le 1$.

\Medskip  Proceeding by
induction on $h\le h_{max}$ and using reactions connecting
classes of height $h-1$ to classes of height $h$, we get the result.  \hfill \eop

%%%%%%%%%%%%%%%%%%%%%%%%%%%%%

\section{(Top) implies  dynamical 
autocatalysis for dilute networks}  \label{section:4}

%%%%%%%%%%%%%%%%%%%%%%%%%%%%%%%%ù

We show here the second part of Theorem \ref{th:main}, and
prove spontaneous autocatalysis (i.e. exponential
amplification of some species starting from an arbitrary
initial condition with low concentrations). 

\Medskip The following notations are used. Reversible 1-1 reactions (for
which some arbitrary orientation is chosen)
are denoted
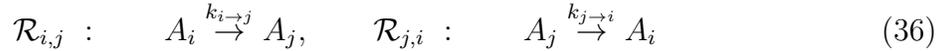
\BEQ {\cal R}_{i,j} \ : \qquad A_i  \overset{k_{i\to j}}{\to} A_j, \qquad {\cal R}_{j,i} \ : \qquad A_j  \overset{k_{j\to i}}{\to} A_i
\EEQ
Forward, irreversible split reactions coming from a reaction
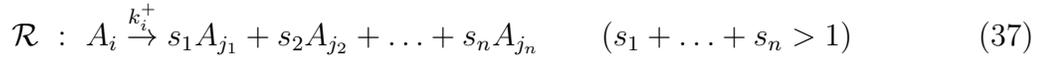
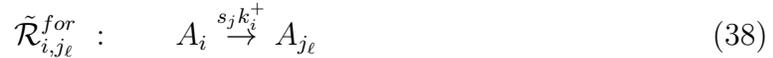
\BEQ {\cal R}\ :\ A_{i}\overset{k_i^+}{\to} s_1 A_{j_1}+s_2 A_{j_2}+\ldots +s_n A_{j_n} \qquad  
(s_1+\ldots+s_n>1)   \label{eq:forward-irreversible}
\EEQ 
are denoted
\BEQ  \tilde{\cal R}^{for}_{i,j_{\ell}}\ : \qquad 
A_i  \overset{s_j k^+_i}{\rightarrow} A_{j_{\ell}}
\EEQ 
Combining 
all these reactions defines (see section \ref{section:3} {\bf A.}) an auxiliary Markov chain $(\tilde{X}(t))_{t\ge 0}$, whose adjoint generator we denote $\tilde{M}$.  On the other hand, the linearized time-evolution
generator of the reaction network containing all reversible 1-1 reactions and forward, irreversible reactions 
({\em excluding} possible degradation reactions) is called $M$. It
is a generalized adjoint Markov generator; we shall use  the path representation of resolvents of $\tilde{M}$ and $M$ introduced
in Suppl. Info. (\S \ref{subsection:GM}).

\Medskip Choose a set of degradation rates $(\alpha_i)_{i\in{\cal S}}>0$ -- we remind the reader that $M$ itself
involves  by assumption no degradation reaction.  Discrete-time transition rates are  
\BEQ w(\alpha)_{i\to j}:=\frac{(M_{\alpha})_{ji}}{|(M_{\alpha})_{i,i}|} = \frac{M_{ji}}{|M_{i,i}|+\alpha_i} \EEQ
for $M_{\alpha}:=M-\alpha$, and similarly
\BEQ \tilde{w}(\alpha)_{i\to j}:=\frac{(\tilde{M}_{\alpha})_{ji}}{|(\tilde{M}_{\alpha})_{i,i}|} 
= \frac{\tilde{M}_{ji}}{|\tilde{M}_{i,i}|+\alpha_i}\EEQ
for $\tilde{M}_{\alpha}:=\tilde{M}-\alpha$.

\Medskip The general purpose of this section is to prove
that  a diluted network satisfying the topological hypothesis (Top) of Theorem
\ref{th:main} is weakly dynamically autocatalytic, provided
it is degradationless, or degradation reactions have
small enough rates.  Furthermore,
we shall be able to prove strong dynamical autocatalysis in some cases, including the
irreducible case.

\Bigskip {\bf A. Irreducible case.} We assume here that the split graph
$G({\mathbb S})$ is irreducible, and prove
strong dynamical autocatalysis. Define $M$ as above (or replace $M$ by $M-\beta$, where $(\beta_i)_{i\in{\cal S}}$
is a set of small enough degradation rates). For any $\alpha\ge 0$,  let $ R(\alpha)$  be its resolvent, with coefficients in $[0,+\infty]$
given by the path representation (\ref{eq:traj}); in Suppl. Info., it
is proved that positivity of the Lyapunov exponent of $M$
  is equivalent to having
\BEQ (R(\alpha))_{i,j}=+\infty  \label{eq:Ralpha=infty} \EEQ
for some (or all) $i,j\in {\cal S}$ and some $\alpha>0$.
Then this condition implies
dynamical autocatalysis for degradation rates $<\alpha$. In
turn, Lemma \ref{lem:PF} and the discussion below
give quantitative criteria for spontaneous autocatalysis.   So let us prove (\ref{eq:Ralpha=infty}).

\Medskip
 By hypothesis, there exists at least one forward irreversible reaction as in (\ref{eq:forward-irreversible}); reindexing, we assume that $i=0$ and $j_{\ell}=\ell,\ell=1,\ldots,n$. Choose a set of degradation rates 
$(\alpha_i)_{i\in{\cal S}}>0$. The generalized adjoint Markov generator $M-\alpha$ and the adjoint sub-Markov generator
 $\tilde{M}-\alpha$ have same off-diagonal coefficients, but diagonal coefficients of $M$ are larger than those of $\tilde{M}$.
 Namely (decomposing $M$ into a  sum of contributions by individual split reactions, see \S \ref{subsection:linearized}), $M({\cal R})=\tilde{M}({\cal R})$ if ${\cal R}$ is
 reversible, while 
 \BEQ \sum_{\ell} M(\tilde{\cal R}^{for}_{i,j_{\ell}})_{i,i}
 =-k_i^+> \sum_{\ell} \tilde{M}(\tilde{\cal R}^{for}_{i,j_{\ell}})_{i,i} = -(s_1+\ldots+s_n)k_i^+ 
 \EEQ
for a forward irreversible reaction.
 Now
\BEA && M_{i,i}=\sum_{{\cal R}\ {\mathrm{reversible}}} M({\cal R})_{i,i}
+ \sum_{\tilde{\cal R}^{for} \ {\mathrm{irreversible}}} 
M(\tilde{\cal R})_{i,i}  \nonumber\\
&&\qquad \ge \sum_{{\cal R}\ {\mathrm{reversible}}} \tilde{M}({\cal R})_{i,i}
+ \sum_{\tilde{\cal R}^{for} \ {\mathrm{irreversible}}} 
\tilde{M}(\tilde{\cal R})_{i,i}    \label{eq:111}
\EEA 
The inequality is strict for $i=0$. It follows: $\tilde{w}(\alpha)_{i\to j}\le w(\alpha)_{i\to j}$, and in particular, $\tilde{w}(\alpha)_{0\to j}<w(\alpha)_{0\to j}$
if $j=1,\ldots,n$.

\Medskip  Then
\BEQ (R(\alpha))_{0,0}=\frac{1}{|(M_{\alpha})_{0,0}|} \sum_{p\ge 0} (f(\alpha)_{0\to 0})^p   \label{eq:112}
\EEQ
where $f(\alpha)_{0\to 0}$ is the total weight of {\em excursions}  from $0$ to $0$, computed using transition rates $w(\alpha)$, namely,
$f(\alpha)_{0\to 0}=\sum_{\ell\ge 1} \sum_{0=x_0\to x_1\to \cdots
\to x_{\ell}\to 0=x_{\ell+1}} \prod_{k=0}^{\ell} w(\alpha)_{x_k\to x_{k+1}}$, where the sum is restricted to paths $(x_k)_{1\le k\le \ell}$  of
length $\ge 1$ in ${\cal S}\setminus\{0\}$. Summing over all possible first steps, we get
\BEQ f(\alpha)_{0\to 0}=\sum_{i\not=0} w(\alpha)_{0\to i} f(\alpha)_{i\to 0},
\EEQ
where $f(\alpha)_{i\to 0}$ is the total weight of paths in
${\cal S}\setminus\{0\}$ issued from $i$, with a final additional step leading back to $0$. In turn, using again  the path representation, we see that $f(\alpha)_{i\to 0}$ may be written as an infinite series whose coefficients are
products of transition rates $w(\alpha)$. 

\Medskip Similarly, one may define $\tilde{f}(\alpha)_{0\to 0}=
\sum_{i\not=0} \tilde{w}(\alpha)_{0\to i} \tilde{f}(\alpha)_{i\to 0}$, where $\tilde{f}(\alpha)_{i\to 0}$ is the same sum as $f(\alpha)_{i\to 0}$, but with  transition rates $w(\alpha)$  replaced by $\tilde{w}(\alpha)$.

\Medskip When $\alpha=0$, $\tilde{f}(0)_{0\to 0}$ is simply the probability for the true (i.e. probability-preserving) Markov chain $\tilde{X}$ to get back to $0$. Irreducible Markov chains
with  finite state space are recurrent, so $ \tilde{f}(0)_{0\to 0}=1$. Now $w(\alpha)\ge \tilde{w}(\alpha)$ (implying $f(\alpha)_{i\to 0}\ge \tilde{f}(\alpha)_{i\to 0})$ and $w(\alpha)_{0\to i}>
\tilde{w}(\alpha)_{0\to i}$, hence (by a simple continuity argument
w. r. to $\alpha$)  $f(\alpha)_{0\to 0}>1$ for $\alpha>0$ small enough, implying
\BEQ (R(\alpha))_{0,0}=+\infty. 
\EEQ

\Bigskip {\bf B. Reducible case.}
We start with a one-parameter family of examples to show the
variety of autocatalytic behaviors (see p. 13)  \label{example:red}.
Let ${\cal C} = \{A_1,A_2\}$ and ${\cal C}'=\{B\}$ be two classes 
with probabilistic flow flowing from ${\cal C}$ into 
${\cal C}'$, 
and (in the $(A_1,A_2,B)$-basis)
$M := \left[\begin{array}{ccc}
-1 & 2 & 0 \\ 1 & -1-k & 0 \\
0 & 1+k & m-1 \end{array}\right]$ 
corresponding to the reaction network
\BEQ A_1\to A_2, \qquad A_2  \to
2A_1 + B, \qquad  B \to mB, \qquad A_2 \overset{k}{\to} B  
\EEQ
with $m > 0$;  all kinetic rates, except for the last
one, are equal to $1$. This network satisfies (Top), hence 
is stoechiometrically autocatalytic,  but its dynamical
status  turns out to depend on the kinetic
rate $k$ of the purely external reaction $A_2\to B$ coupling
  ${\cal C}$ to ${\cal C}'$. Namely,   
 the determinant  of  $ M\Big|_{{\cal C}}=\left[\begin{array}{cc} -1 & 2 \\ 1 & -1 -k
\end{array}\right]$  is $k-1$, and its trace is $<0$, implying
that both its eigenvalues are $<0$ if the coupling constant
$k$ is $>1$. If, furthermore, $m=1$ (i.e. $M\Big|_{{\cal C}'}$ 
is not autocatalytic), then the Lyapunov exponent is $0$.
This is easily understandable:  the purely external reaction $A_2\to B$ acts 
as a degradation reaction for the system restricted to the
minimal class $\cal C$, and $k$ as a degradation rate.

\Medskip
Assuming a weaker degradation rate ($k<1$), the Lyapunov
exponent becomes $>0$. To keep computations simple, we
simply let $k=0$ (no degradation rate). Then the Lyapunov
exponent of $M\Big|_{{\cal C}}$  
is $\lambda_{max} :=\sqrt{2}-1$, and
$v_{\cal C} :=
\left[\begin{array}{c} \sqrt{2} \\ 
1\end{array}\right]$ 
is an associated positive eigenvector; that of $M\Big|_{{\cal C}'}= \, [ m-1 \, ]$ is $m-1$.
The maximum (Lyapunov) eigenvalue of $M$ is $\max(\lambda_{max},m-1)$. There are three cases,
depending on the spectral parameter $m$:
\begin{enumerate}
\item if $\lambda_{max} > m - 1$, then $v_{\cal C}$ can be extended into a positive Lyapunov eigenvector for
$M$, implying strong dynamical autocatalysis;
\item if $\lambda_{max} = m - 1$ (resonant case), then this is not possible (the associated Jordan
block is not trivial). Instead, one gets the downstream Lyapunov eigenvector $\left[\begin{array}{c}
 0 \\ 0 \\ 1 \end{array}\right]$. 
Thus dynamical autocatalysis holds only in the weak sense;

\item  if $\lambda_{max} < m - 1$, then $\left[\begin{array}{c}
 0 \\ 0 \\ 1 \end{array}\right]$
 is again a downstream Lyapunov eigenvector, and
dynamical autocatalysis holds only in the weak sense.

\end{enumerate}

\Medskip  Our proof of weak autocatalysis encompasses all cases without addressing such spectral
considerations. It follows from our argument in {\bf A.} through an elementary perturbation
argument. Namely, replace the above matrix $M$ by 
$ M(\eta,\eps) := M(\eta) + \eps J$, where $\eps > 0$ is a
small parameter,  $J$ is an off-diagonal matrix with non-negative coefficients,  and 
$M(\eta):= M_+ + \eta M_{ext}$, where $\eta M_{ext}$ is the sum of the generators
associated to purely external reactions; the
parameter $\eta>0$ determines the order of magnitude of the
coupling between classes induced by purely external reactions. If $J$ has
enough nonzero coefficients, then $M(\eta,\eps)$ will be irreducible. (Assuming all concentrations of all species
are $> 0$, this may e.g. be achieved by including also some split  reverse reactions  coming from
 reverse reactions $s_1 A_{j_1}+s_2 A_{j_2}+\ldots +s_n A_{j_n} \overset{k_i^-}{\to}
A_i$ connecting classes upwards, 

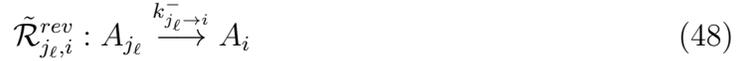
\BEQ \tilde{\cal R}^{rev}_{j_{\ell},i}: A_{j_{\ell}} \overset{k^-_{j_{\ell}\to i}}{\longrightarrow} A_i
\EEQ
with $k^-_{j_{\ell}\to i}= k_i^- s_{j_{\ell}} [A_{j_{\ell}}]^{s_{\ell}-1}
\ \prod_{\ell'\not=\ell} [A_{j_{\ell'}}]^{s_{\ell'}}$, see
\S \ref{subsection:linearized}). 

Now $M(\eta,\eps)$ is an irreducible generalized
Markov matrix. The Perron-Frobenius theorem implies that $\lambda_{max}(\eta,\eps) := \lambda_{max}(M(\eta,\eps))$ has
multiplicity $1$, and that there exists a unique associated eigenvector  $v(\eta,\eps) = (v_i(\eta,\eps))_{i\in {\cal S}}$
such that $v_i(\eta,\eps) > 0$ for all $i$, and
$\sum_{i\in {\cal S}} v_i(\eta,\eps) = 1$. Following the arguments in {\bf A.} for $M_+$,  one sees
that the addition of $\eta M_{ext}$ and  $\eps J$ modifies transition rates $w(\alpha)$ only by $O(\eta)+ O(\eps)$, thus $\lambda_{max}(\eta,\eps) \ge \lambda > 0$ for some constant $\lambda$ uniformly in $\eta,\eps$ if $\eta,\eps$ are small enough. Following the compacity
argument in (Stewart 1990, Th. 6.10.),  and assuming
$\eta$ to be small enough,  one proves the existence of a limiting eigenvector $v(\eta)$ such that

\BEQ M(\eta)v(\eta) = \lambda_{max}(\eta)v(\eta),
\EEQ

 where $v(\eta)$ is the limit of some subsequence $(v(\eta,\eps_k))_{k=1,2,\ldots}$  with $\eps_k\to 0$, and

\BEQ \lambda_{max}(\eta) := \lim_{\eps\to 0} \lambda_{max}(\eta,\eps) \ge \lambda > 0
\EEQ
 By continuity,
$\sum_{i\in {\cal S}} v_i(\eta)= 1$ so that $v(\eta)$ is normalized, but
coefficients $v_i(\eta), i \in {\cal S}$ are only non-negative in general. Thus we have shown the following:  {\em  the network with generator $M(\eta)$
is weakly dynamically
autocatalytic   with Lyapunov eigenvalue $\lambda_{max}(\eta)$, provided the coupling coefficient $\eta$ is small enough.}

\Bigskip Let us briefy discuss specific hypotheses (generalizing case 1. above) under which strong
dynamical autocatalysis holds (for $\eta$ small enough). First, the Perron block decomposition (see (Stewart 1990)) of $M$
implies that its spectrum $\Sigma(M)$ is the union of the spectra  $ (\Sigma(M\Big|_{\cal C}))_{\cal C}$  of its restrictions
to all classes $\cal C$. We assume (i) that there exists a minimal class $\cal C$ from which all classes
can be attained from it following the probabilistic 
flow (i.e. following arrows downward);
(ii) that $\lambda_{max}(M\Big|_{\cal C})=
\lambda_{max} := \max(\Sigma(M)){\color{red}>0}$ is the maximum of all Lyapunov exponents
of all classes, more precisely,
$\lambda_{max}(M\Big|_{{\cal C}'}) < \lambda_{max} $ for
  all 
${\cal C}'\not={\cal C}$. By the above compacity
argument, there exists a nonzero eigenvector $v\ge 0$  such that $Mv = \lambda_{max}v$. Restricting
to $\cal C$, we get a positive Lyapunov eigenvector $v_{\cal C} = (v_i)_{i\in {\cal C}} > 0$ for $M\Big|_{\cal C}$. Hypothesis (i)
then implies that $e^{tM}v > 0$ for all $t > 0$. Now $e^{tM}v = e^{t\lambda_{max}}v$, so that $v > 0$ is positive,
and strong autocatalysis is proven. To be concrete (as a final remark), exponential growth
(76) will hold for all components for time values
$ t > \tau$ , where $\tau > 0$ is a homogeneization
time as in Lemma \ref{lem:PF} (ii) (see discussion below the Lemma).

%%%%%%%%%%%%%%%%%%%%%%%%%%%%
%%%%%%%%%%%%%%%%%%%%%%%%%%%%%%%ù

\section{Perspectives}

%%%%%%%%%%%%%%%%%%%%%%%%%%
%%%%%%%%%%%%%%%%%%%%%%%%

\Medskip We have introduced in our main result, Theorem \ref{th:main}, a condition (Top) that 
provides a topological characterization of
autocatalysis in the dilute regime (i.e. for low concentrations).
This characterization is complete in the limit of negligible degradation rates: indeed, in this case, (Top) is necessary and sufficient for autocatalysis both in the stoechiometric and in the dynamical sense (at least, for irreducible networks).
We have furthermore shown that, in this limit, an infinitesimal amount of any species participating in the autocatalytic network ensures the onset of dynamical autocatalysis (see Lemma \ref{lem:PF}).
In practice, this means that autocatalytic amplification can start spontaneously upon the rare appearance of a single autocatalyst.
Interestingly, these conclusions directly apply to the universal minimal autocatalytic networks (autocatalytic cores) found in any autocatalytic system  (Blokhuis 2020), as they all respect (Top).

\Medskip A first excursion out of this well-understood regime consists in including significant degradation reactions. Then condition (Top) remains necessary and sufficient for stoechiometric autocatalysis, but is only necessary for dynamical autocatalysis.
Determining  viability thresholds, i.e. maximum combinations of degradation rates which allow dynamical autocatalysis, is critical for the design of  autocatalytic reaction networks and in origin of life studies (Jeancolas 2020).
Using branching processes, viability thresholds were determined for autocatalytic cores in the stochastic regime where only a few molecules are present  (Blokhuis 2020).
Specifically, it was shown there that a {\em single} molecule survives with positive probability if and only if a certain inequality involving kinetic and degradation rates is satisfied.
A next step of the treatment presented here will be to characterize viability thresholds allowing positivity of the Lyapunov exponent, and understand the relationship between the continuous and stochastic treatments of the viability thresholds.
Notably, a conclusion of the stochastic treatment is that a multiplicity of internal catalytic cycles within the autocatalytic network favors survival (equivalently, allows larger degradation rates). It is tempting to speculate that this conclusion should apply as well to viability thresholds in the kinetic limit, as derived from the study of Lyapunov exponents.

\Medskip Another direction for generalization is to go beyond the diluted regime.
Away from it,  Lyapunov exponents characterize stability in the neighborhood of stationary points other than  the zero concentration limit, 
including equilibrium for networks satisfying detailed balance 
and growth modes for systems with dilution rate ensuring constant total concentration (as in (Eigen 1979)). 
However, in all generality, there is  not necessarily a direct relationship between positivity of the Lyapunov exponent (growth rate) of the {\em linearized} system and the growth of the original nonlinear dynamical system. 

\Bigskip In a companion paper (Unterberger 2021), we
discuss all these points using an approach based on the
analysis of \S \ref{subsection:linearized} and \S \ref{subsection:GM}. 
We obtain  the following tentative conclusions, valid in the non-diluted regime:
(i) Topology {\em and} thermodynamics together inform about autocatalysis; 
(ii) Estimating the Lyapunov exponent is (despite objections
raised in the previous paragraph) a useful 'proxy' allowing quantitative estimates of  the growth rate.
General quantitative statements include: the
computation of  certain autocatalytic thresholds in the diluted regime; and  
 estimates for Lyapunov exponents depending essentially on the topology of the network and on 
thermodynamics {\em for arbitrary concentrations}. We also show on examples that the curves of 'proxy' dynamical systems based on the above estimates compare well to the curves obtained by numerical integration, over a surprisingly large range of growth regimes.

\Bigskip The approach developed here and in our companion paper is a promising one for the  investigation  of more complex networks. Indeed, it shows that partial knowledge based on topology and thermodynamics informs on dynamics, independently of the knowledge of reaction rate constants, which is generally missing.
A particularly important question is to understand the conditions for the existence of  multiple growth modes that could support rudimentary forms of Darwinian evolution (Fernando 2011). Together with  threshold estimates, this may allow us to build scenarios for the emergence of evolution during the origin of life (Jeancolas 2020).

%%%%%%%%%%%%%%%%%%%%%%
%%%%%%%%%%%%%%%%%%%%%%%

\section{Supplementary information}

%%%%%%%%%%%%%%%%

For simplicity, the reaction networks we consider in this appendix
contain only {\em forward} reactions with {\em one}
reactant,
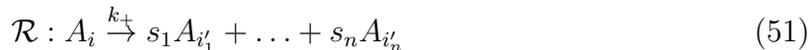
\BEQ {\cal R}: A_i \overset{k_+}{\to}  s_1 A_{i'_1}+\ldots+s_{n} A_{i'_{n}}  \label{eq:forward-reaction}
\EEQ
and the corresponding {\em reverse} reactions,
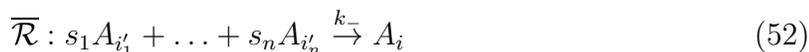
\BEQ \overline{\cal R}: s_1 A_{i'_1}+\ldots+s_{n} A_{i'_{n}} \overset{k_-}{\to} A_i \label{eq:reverse-reaction}
\EEQ
All reaction networks discussed in the  article are of this type.

\Medskip
This section is organized as follows. We start by presenting Type I and Type III cycles; contrary to  (Blokhuis 2020), we assume stoechiometry 1 for reactants, in conformity with (\ref{eq:forward-reaction}). Extended results for these cycles are presented
in the companion paper (Unterberger 2021). A short
argument for Theorem \ref{th:main} (1) (characterization
 of stoechiometric autocatalysis) is presented in 
 \S \ref{SI:Perron} in the case of an irreducible
 network.  Then (\S \ref{subsection:linearized}), we write down explicit formulas for the linearized time-evolution generator $M$ of a reaction network.
Finally, we present in \S \ref{subsection:GM} a "theory in a nutshell" for
generalized adjoint Markov generators, extending results known from
Markov chain theory; see in particular Lemma \ref{lem:resolvent} and \ref{lem:PF}.

%%%%%%%%%%%%%%%

\subsection{Presentation of type I cycles}

%%%%%%%%%%%%%%%%%%%%%%%%%%%%%%

We  consider in this subsection type I cycles of arbitrary length in the Blokhuis-Lacoste-Nghe
classification,

\bigskip

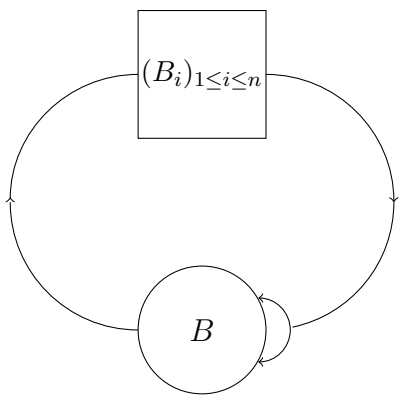
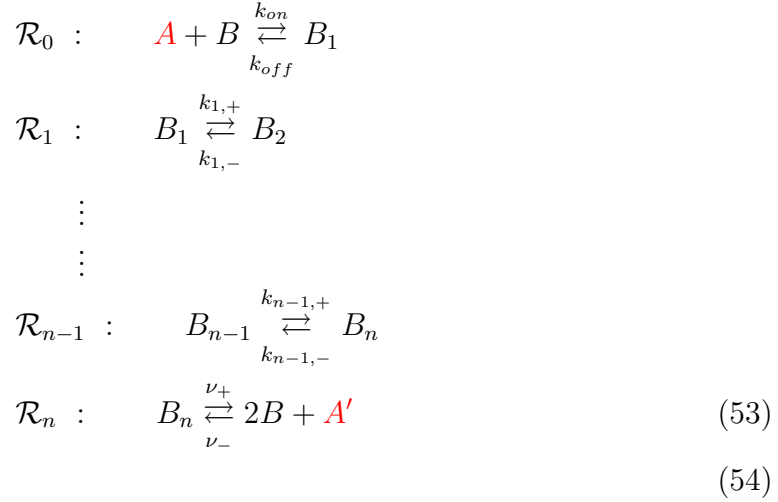
{\centerline{
\begin{tikzpicture}[scale=0.85]
\draw(-1,-1) rectangle(1,1);  \draw(0,0) node {\small $(B_i)_{1\le i\le n}$};
\draw[->](1,0) arc(90:0:2 and 2);
%\draw(2.8,-2-0.2) rectangle (3.2,-2+0.2); 
\draw[-<](-1,0) arc(90:180:2 and 2);
%\draw(-3.2,-2-0.2) rectangle(-2.8,-2+0.2);
\draw(3,-2) arc(0:-78:2 and 2);
\draw(0,-4) circle(1);  \draw(0,-4) node {$B$};
\draw(-1,-4) arc(-90:-180:2 and 2);
\draw[<->](0.88,-3.5) arc(90:-90:0.5 and 0.5);
\end{tikzpicture}
}}

\BEA && {\cal R}_0\ : \qquad  {\color{red} A} + B\overset{k_{on}}{\underset{k_{off}}{\rightleftarrows}} B_1 \nonumber\\
&&  {\cal R}_1\ : \qquad B_1 \overset{k_{1,+}}{\underset{k_{1,-}}{\rightleftarrows}} B_2
\nonumber\\
&&\qquad\vdots \nonumber\\  &&\qquad\vdots \nonumber\\ 
&& {\cal R}_{n-1}\ : \qquad B_{n-1} \overset{k_{n-1,+}}{\underset{k_{n-1,-}}{\rightleftarrows}} B_n
\nonumber\\
&& {\cal R}_n\ :\qquad  B_n \overset{\nu_+}{\underset{\nu_-}{\rightleftarrows}}
2B+ {\color{red} A'}  \\
\EEA

\Medskip The simple model studied in section 2 corresponds to
the special case $n=1$. Chemostatted species -- to be thought of
e.g. as redox/energy carrier couple as in section 2 -- are $(A,A')$
 (in red). Dynamical  species are $(B,B_1,\ldots,B_n)$. Leaving out
chemostatted species, we have a cycle $B\leftrightarrows B_1 \leftrightarrows B_2\leftrightarrows
\cdots\leftrightarrows B_n\leftrightarrows 2B$. 

\Medskip We have assumed trivial $1\leftrightarrow 1 $  stoechiometry for all reactions along the cycle, except
for the {\em duplication reaction} $({\cal R}_n): 
B_n\longrightarrow 2B$ closing the cycle. There is nothing special about stoechiometry 2. The extension to the case when $({\cal R}_n) : B_n\longrightarrow mB$ with arbitrary $m=2,3,\ldots$ is straightforward. 

\Medskip 
Type I cycles
are autocatalytic in the stoechiometric sense, as seen
by choosing any reaction vector $c=(c_0,\ldots,c_n)$ such that
$c_0>c_1>\ldots>c_n>c_0/2>0$.  On the other hand, choosing
$c=(1,\ldots,1)$ yields the coarse-grained duplication reaction 
for species $B$
\BEQ {\color{red} A} + B \to
2B+ {\color{red} A'}.\EEQ

%%%%%%%%%%%%%%%%%%%%%%%%%

\subsection{Presentation of type III cycles}
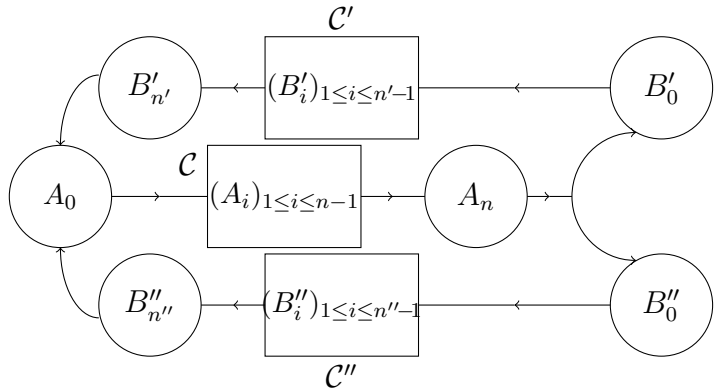
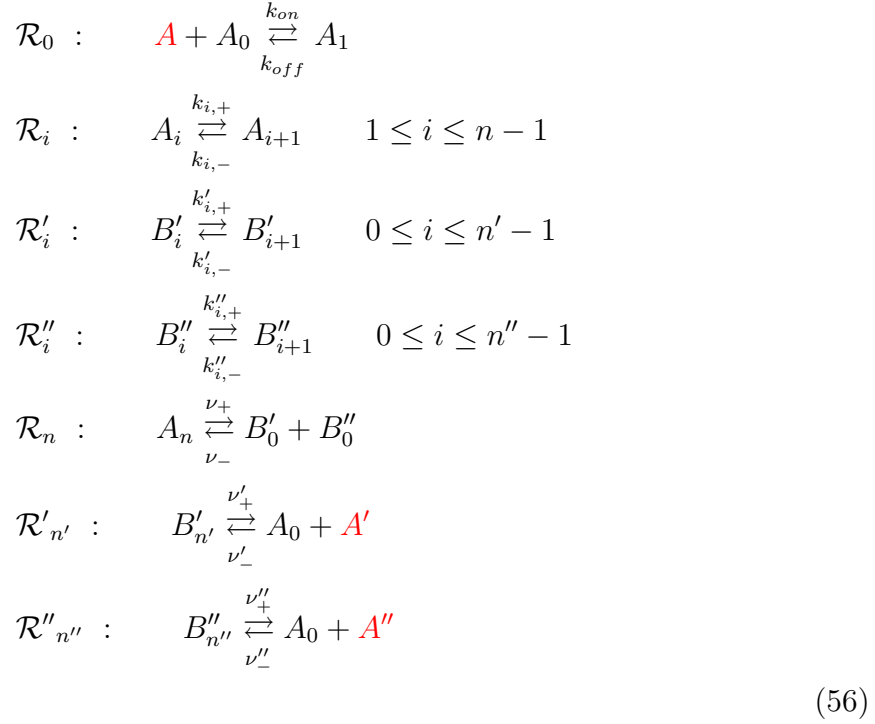
\label{SI:typeIII}

%%%%%%%%%%%%%%%%%%%%%%%%%%%%%%

{\centerline{
\begin{tikzpicture}[scale=0.85]
\draw(0,0) circle(0.8);  \draw(0,0) node {\small $A_0$};
\draw[->](0.8,0)--(1.55,0); \draw(1.55,0)--(2.3,0); 
\draw(2.3,-0.8) rectangle(4.7,0.8);  
\draw(2,0.5) node{\bf ${\cal C}$};
\draw(3.5,0) node {\small $(A_i)_{1\le i\le n-1}$};
\draw[->](4.7,0)--(5.2,0); \draw(5.2,0)--(5.7,0);
\draw(6.5,0) circle(0.8); \draw(6.5,0) node {$A_n$};
\draw[->](7.3,0)--(7.65,0); \draw(7.65,0)--(8,0);
\draw[<->](9,1) arc(90:270:1 and 1);
\draw(9.4,1.7) circle(0.8); \draw(9.4,1.7) node {$B'_0$};
\draw(9.4,-1.7) circle(0.8); \draw(9.4,-1.7) node {$B''_0$};
%%%%%%%%%%%%%%
\draw[->](8.6,1.7)--(7.1,1.7); \draw(7.1,1.7)--(5.6,1.7); \draw(5.6,2.5) rectangle(3.2,0.9);
\draw(4.4,2.8) node {${\cal C}'$};
\draw(4.4,1.7) node {\small $(B'_i)_{1\le i\le n'\!-\!1}$};
\draw[->](3.2,1.7)--(2.7,1.7); \draw(2.7,1.7)--(2.2,1.7);  
\draw(1.4,1.7) circle(0.8); \draw(1.4,1.7) node {$B'_{n'}$};
\draw[<-](0,0.8) arc(180:90:0.6 and 1.1);
%%%%%%
\draw[->](8.6,-1.7)--(7.1,-1.7); \draw(7.1,-1.7)--(5.6,-1.7); \draw(5.6,-2.5) rectangle(3.2,-0.9);
\draw(4.4,-2.8) node {${\cal C}''$};
\draw(4.4,-1.7) node {\small $(B''_i)_{1\le i\le n''\!-\!1}$};
\draw[->](3.2,-1.7)--(2.7,-1.7); \draw(2.7,-1.7)--(2.2,-1.7);  
\draw(1.4,-1.7) circle(0.8); \draw(1.4,-1.7) node {$B''_{n''}$};
\draw[<-](0,-0.8) arc(-180:-90:0.6 and 1.1);
%%%%%%%%%%
%%%%%%%%%%%%
\end{tikzpicture}
}}

\bigskip

\BEA && {\cal R}_0\ : \qquad  {\color{red} A} + A_0\overset{k_{on}}{\underset{k_{off}}{\rightleftarrows}} A_1 \nonumber\\
&&  {\cal R}_i\ : \qquad A_i \overset{k_{i,+}}{\underset{k_{i,-}}{\rightleftarrows}} A_{i+1}
\qquad 1\le i\le n-1
\nonumber\\
&& {\cal R}'_i\ : \qquad B'_i \overset{k'_{i,+}}{\underset{k'_{i,-}}{\rightleftarrows}} B'_{i+1}
\qquad 0\le i\le n'-1
\nonumber\\
&& {\cal R}''_i\ : \qquad B''_i \overset{k''_{i,+}}{\underset{k''_{i,-}}{\rightleftarrows}} B''_{i+1}
\qquad 0\le i\le n''-1 \nonumber\\
&& {\cal R}_n\ : \qquad A_n \overset{\nu_+}{\underset{\nu_-}{\rightleftarrows}} B'_0 + B''_0 
\nonumber\\
&& {\cal R'}_{n'}\ : \qquad B'_{n'} \overset{\nu'_+}{\underset{\nu'_-}{\rightleftarrows}} A_0 + {\color{red} A'}
\nonumber\\
&& {\cal R''}_{n''}\ : \qquad B''_{n''} \overset{\nu''_+}{\underset{\nu''_-}{\rightleftarrows}} A_0 + {\color{red} A''}
\nonumber\\
\EEA

\Medskip Chemostatted species (in red) are $(A,A',A'')$. Dynamical  species are $(A_i)_{0\le i\le n},(B'_i)_{0\le i\le n'}$, \\
$(B''_i)_{0\le i\le n''} $. 

\Medskip We have  trivial $1\leftrightarrow 1 $  stoechiometry for all reactions along the two cycles, 
and $1\to 1+1$ for the {\em pitchfork reaction} $A_n\to B'_0+B''_0$.
Choosing a positive reaction vector such that $c_0>\ldots>c_n$, $c'_0>\ldots>c'_{n'}$, $c''_0>\ldots>c''_{n''}$, and 
$c_n>\max(c'_0,c''_0)$, $c'_{n'}+c''_{n''}>c_0$, one obtains
a positive balance for all species. Choosing instead $c=(1,\ldots,1)$ yields the coarse-grained duplication reaction for species $A_0$
\BEQ {\color{red} A} + A_0 \to 2A_0 + {\color{red} A'+A''}.
\EEQ

%%%%%%%%%%%%%%%%%%%%%%%

\subsection{Autocatalysis from the stoechiometric matrix}  \label{SI:Perron}

We give here a short argument for Theorem 
\ref{th:main} (1)  in the irreducible case.

\Medskip
We consider an irreducible component of the reaction network with $n$ species verifying (Top): every reaction has  exactly one reactant and at least one reaction has $\ge 2$ products or a product with a stoechiometry strictly $\ge 2$.
Correspondingly, each column $j$ of the stoechiometric matrix ${\mathbb S}$ possesses a coefficient $s_{ij}=-1$  and otherwise positive coefficients such that 
\BEQ \sum_{i=1}^n s_{ij} \geq 0. \label{eq:s} \EEQ Additionally, there is a column index $k$ such that  $\sum_{i=1}^n s_{ik} > 0$.

\Medskip 
We want to show that there exists a reaction vector $c>0$ such that ${\mathbb S}c>0$, i.e.   $({\mathbb S}c)_i>0$ for all $i$. For this, it is sufficient to show that $Mc'>0$ for a certain reaction vector $c'>0$, where $M$ is a matrix whose columns are positive linear combinations of those of ${\mathbb S}$, as constructed below.

\Medskip
Let $J(i)=\{j\ |\  s_{ij} = -1\}$ be the set
of reactions having species $i$ as reactant,  and  $N_i$  the cardinal of $J(i)$. 
As the network is irreducible, $N_i \geq 1$ for all $i$.
Let $n:=|{\cal S}|$. 
Denoting $C_M^j$, resp. $C_S^j$ the $j$-th column of $M$, resp.  ${\mathbb S}$,
we let, for $j=1,...,n$:
$$C_M^j := \frac{1}{N_j} \sum_{j' \in J(j)} C_{\mathbb S}^{j'}.$$

\Medskip
By construction, $M \equiv A-I$, where $A$ is square and non-negative, i.e. $A_{ij}\ge 0$ for all $i,j$. Given that the network is strongly connected, $A$ is irreducible. Stoechiometric hypotheses (\ref{eq:s}) impose $\sum_{i=1}^n A_{ij} \geq 1$ for every $j$ and  $\sum_{i=1}^n A_{ik} > 1$ for a certain $k$. 
By the Perron-Frobenius theorem, the largest eigenvalue $\lambda$ of $A$ is positive and associated with an  eigenvector $c'>0$.
We have:
$$\lambda \sum_{i=1}^n c'_i = \sum_{i=1}^n (\sum_{j=1}^n A_{ij} c'_j) = \sum_{j=1}^n (\sum_{i=1}^n A_{ij}) c'_j > \sum_{j=1}^n c'_j$$

\Medskip
This implies $\lambda>1$. Hence $c'$ is a positive eigenvector
of $M=A-I$ with eigenvalue $\lambda-1>0$.

%%%%%%%%%%%%%%%%%%%%%%%%%%%%%%%%%%%%%%%%%%%

\subsection{Linearized time-evolution
generator for reaction networks}  \label{subsection:linearized}

%%%%%%%%%%%%%%%%%%%%%%%%%%

The linearized time-evolution
generator $M=M([A])$ of a reaction network has been defined in eq. (\ref{eq:3})--(\ref{eq:5}). The  current $J_{{\cal R}}=k_+[A_i]$
associated to a forward reaction ${\cal R}$ as in (\ref{eq:forward-reaction}) is straightforwardly linearized to $J_{lin,{\cal R}}(A)=k_+ A_i$.  Considering now a reverse reaction (\ref{eq:reverse-reaction}), the  reverse current is 
$k_- \prod_{\ell=1}^n [A_{i'_{\ell}}]^{s_{\ell}}$, yielding
a  linearized current 
\BEQ J_{lin}:=\sum_{\ell} J_{lin}^{\ell} A_{i'_{\ell}}, \qquad 
J_{lin}^{\ell}:= k_- s_{\ell} \Big(\prod_{\ell'\not=\ell}   [A_{i'_{\ell'}}^{s_{\ell'}}]
\Big) [ A_{i'_{\ell}}]^{s_{\ell}-1}.
\label{eq:Jlin}
\EEQ
  The coefficients of the matrix $M$ 
are obtained by summing individual matrices $M({\cal R})$
associated to  linearized  forward reactions ${\cal R}: A_i \overset{k_+}{\to}  s_1 A_{i'_1}+\ldots+s_{n} A_{i'_{n}}$, 
\BEQ \frac{dA_i}{dt}= -k_+ A_i; \qquad 
\frac{dA_{i'_{\ell}}}{dt}=s_{\ell} k_+ A_{i},
\ \ \ell=1,\ldots,n
\EEQ
and  matrices $\sum_{\ell} M(\bar{\cal R},\ell)$ associated to  linearized  reverse reactions $\bar{\cal R}$ (see (\ref{eq:Jlin}))
\BEQ \frac{dA_i}{dt}= J_{lin}^{\ell} A_{i'_{\ell}}; \qquad  \frac{dA_{i'_{\ell'}}}{dt}=-s_{\ell'} J_{lin}^{\ell} A_{i'_{\ell}}, \ \ \ell'=1,\ldots,n.
\EEQ

\Medskip If $\cal R$ is a forward reaction, the corresponding contribution $M({\cal R})$ to $M$ is 
(see below (\ref{eq:5})) a generalized adjoint Markov generator with {\em negative} killing rate $a_i({\cal R})=k_+(1-\sum_{\ell} s_{\ell})$, which vanishes precisely in the
case of a reversible reaction $A_i\leftrightarrows A_j$.

\Medskip Consider now a {\em reverse} reaction $\bar{\cal R}$. The matrix  $M(\bar{\cal R},\ell)$ is {\em not} a generalized adjoint Markov generator if $n\ge 2$, because
of the {\em probability leak currents} $-s_{\ell'}J^{\ell}_{lin}A_{i'_{\ell}}=(M(\bar{\cal R},\ell))_{i'_{\ell'},i'_{\ell}}$ from state $i'_{\ell}\not=i'_{\ell'}$; also, it features
$\ge 0$   killing rates $a_{i'_{\ell}}(\bar{\cal R},\ell)= J^{\ell}_{lin}  \Big[s_{\ell}-1\Big] $, computed  without considering probability
leak currents, considered as  external non-diagonal terms without
probabilistic interpretation. The reverse reaction $\bar{\cal R}\ :\ B''_0+B'_0\overset{\nu_-}{\longrightarrow} A_n$ in type III cores has $n=2$, and  does exhibit
leak currents. Matrices
$M(\bar{\cal R},0')$, resp. $M(\bar{\cal R},0'')$, 
are identified with the two columns of the matrix

\medskip

\begin{tikzpicture}
\draw(0,0) node{$ M(\bar{\cal R}):= M(\bar{\cal R},0') + M(\bar{\cal R},0'')=$};
\draw(4.5,1.5)--(4,1.5)--(4,-1.5)--(4.5,-1.5);
\draw(4.5,1) node{$0$};
\draw(6,1) node {\small $\nu_- [B''_0]$};
\draw(8,1) node {\small $\nu_-[B'_0]$};

\draw(4.5,0) node{\small $0$};
\draw(6,0) node {\small $-\nu_- [B''_0]$};
\draw[red](8,0) node {\small $-\nu_-[B'_0]$};

\draw(4.5,-1) node{\small $0$};
\draw[red](6,-1) node {\small $-\nu_- [B''_0]$};
\draw(8,-1) node {\small $-\nu_-[B'_0]$};

\draw(8.5,1.5)--(9,1.5)--(9,1.5)--(9,-1.5)--(8.5,-1.5);
\label{eq:-}
\end{tikzpicture}

\Medskip  Reverse reactions putting into contact 
$n\ge 2$ different species produce negative off-diagonal coefficients, here emphasized in red. If the resulting matrix $M=\sum_{{\cal R}} M({\cal R}) +\sum_{\bar{\cal R},\ell} M(\bar{\cal R},\ell)$ has negative off-diagonal coefficients, it cannot be interpreted as a generalized Markov generator, therefore Lemma \ref{lem:resolvent}  below (allowing easy upper bounds for the Lyapunov exponent) does not hold.

%%%%%%%%%%%%%%%%%%

\subsection{Generalized Markov generators}  \label{subsection:GM}

%%%%%%%%%%%%%%%%ù
A central notion
in this article is that of {\em generalized Markov processes},
i.e. discrete- or continuous-time Markov processes which are
not necessarily probability-preserving; a general
introductory reference is (Norris 1997, chap. 2). Let ${\cal S}=\{1,\ldots,n\}$ be a finite state space. Then
 an $n\times n$ matrix $M$ is a  {\em generalized adjoint Markov generator} if 
 \begin{itemize}
 \item[(i)] diagonal coefficients $M_{i,i}$, $1\le i\le n$ are $<0$;
 \item[(ii)]  and off-diagonal coefficients
$M_{i,j}$, $i\not=j$ are $\ge 0$.
\end{itemize}
If $\sum_{i=1}^n M_{i,j}=0$ for all $j$, then  coefficients of the transposed matrix $M^t$ sum up to zero on each line,
so that $M^t$ is a conventional probability-preserving Markov generator: 
${\bf 1}^t=\left(\begin{array}{ccc} 1 & \cdots  1
\end{array}\right)$ is a left-eigenvector of $M$ with
eigenvalue $0$. The probability measure $ \mu(t)$ of the process
at time $t$ is $ e^{tM}\mu(0)$, solution of the master equation
$\frac{d}{dt}\mu=M\mu$; probability preservation means
that $\sum_i \mu_i(t)=1$ for all time.  Assume, more generally, that
$a_{j}:=|M_{j,j}|- \sum_{i\not=j} M_{i,j}\ge 0$, then the total
probability 
$\sum_i \mu_i(t)$ is a decreasing function of time, and 
$a_j$ can be interpreted as a killing rate (biologically, a degradation rate) in state $j$. For a finite set of states, there is no obstacle in considering the
case when  killing rates $a_j$ can have either sign. 
We discuss the associated random process $(X(t))_{t\ge 0}$ later on; by definition $X(t)\in {\cal S}$ has transition rate $M_{i,j}$ from $j$ to $i$. By construction,
$\mu_i(t)\equiv \proba[X(t)=i]=(e^{tM}\mu(0))_i$, generalizing
the above master equation, where $\P$ is a (non-normalized)
measure on trajectories.  Probabilistic tools give an intuitive access to the resolvent in terms of trajectories
of the Markov process, from which we derive a characterization and  properties
of the Lyapunov exponent.

\Bigskip {\bf Communicating classes, irreducibility. Example of the
"$A_1 A_2 A_3 \longrightarrow B_1 B_2 B_3$" autocatalytic  reaction network.}
 Let $M$ be a generalized adjoint Markov generator on   ${\cal S}=\{1,\ldots,n\}$. The matrix $M$ defines a  graph $G(M)$ with vertex
state $\cal S$ and oriented edge set  ${\cal E}(M)$: a pair
$e=(x\to y)$, $x\not=y$  is an edge if $M_{y,x}>0$; the
probability flow follows edges.
Following standard terminology in Markov chains, we say
that $x\not=y$ communicate (which we denote $x\sim y$)
if there exists a path from $x$ to $y$ and a path from $y$ to $x$, namely, a chain of edges $(x\to x_1),(x_1\to x_2),\ldots,
(x_{n}\to y)$ and a chain of edges $(y\to \tilde{x}_1), 
(\tilde{x}_1 \to \tilde{x}_2),\ldots,(\tilde{x}_{n'} \to x)$ 
with $n,n'\ge 0$. Letting also $x\sim x$ for all $x$, this 
defines equivalence classes called {\em communicating classes}. 
$M$ is said to be {\em irreducible} if there is only one class. 

\noindent $M$ is
clearly reducible if the graph is not connected, but this
means that we are dealing with several independent systems, an
uninteresting situation. We may assume instead that the graph
$G(M)$ is always connected. On the other hand, there exist
connected graphs which are not irreducible, for instance the
graph of the  "$A_1 A_2 A_3 \longrightarrow B_1 B_2 B_3$" autocatalytic  reaction network (also discussed in section 
\ref{section:3}), a graph on the set ${\cal S}=\{1,2,3,1',2',3'\}$, 

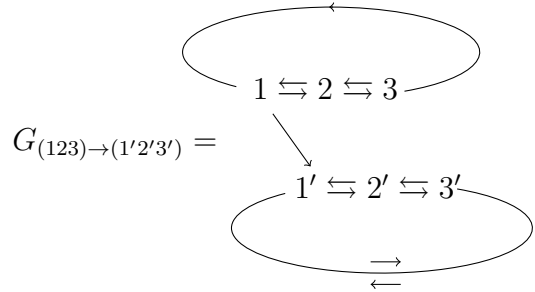
\bigskip
{\centerline{
\begin{tikzpicture}
\draw(-2.8,-0.7) node {$G_{(123)\to (1'2'3')}=$};
\draw(0,0) node{$1 \leftrightarrows 2 \leftrightarrows 3$};
\draw[->](1.05,0) arc(-60:90:2 and 0.6);
\draw(0.1,1.12) arc(90:230:2 and 0.6); 
\draw[->](-0.7,-0.3)--(-0.2,-1);
\draw(0.7,-1.3) node{$1' \leftrightarrows 2' \leftrightarrows 3'$}; 
\draw(1.05+0.7,0-1.3) arc(60:-230:2 and 0.6);
\draw[->](0.7+0.1-0.2,-1.3-1.12+0.15)--(0.7+0.1+0.2,-1.3-1.12+0.15);
\draw[<-](0.7+0.1-0.2,-1.3-1.12-0.15)--(0.7+0.1+0.2,-1.3-1.12-0.15);
\end{tikzpicture}
}}

\Bigskip 
The associated reaction network is

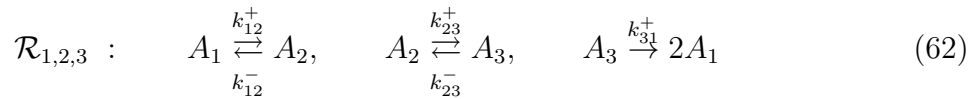
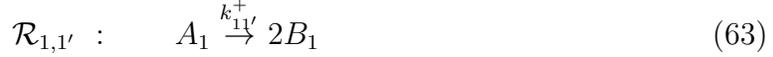
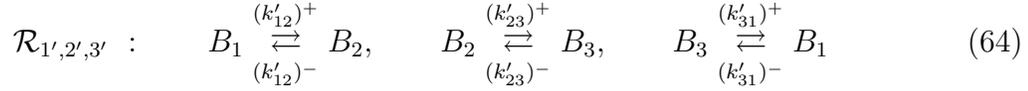
\BEQ {\cal R}_{1,2,3}\ :\qquad  A_1 \overset{k_{12}^+}{\underset{k_{12}^-}{\rightleftarrows}}  A_2, \qquad A_2 \overset{k_{23}^+}{\underset{k_{23}^-}{\rightleftarrows}} A_3, \qquad A_3 \overset{k_{31}^+}{\rightarrow} 2A_1  \EEQ
\BEQ {\cal R}_{1,1'} \ :\qquad  A_1 \overset{k_{11'}^+}{\to}  2B_1 \EEQ
\BEQ {\cal R}_{1',2',3'}\ :\qquad  B_1
\overset{(k'_{12})^+}{\underset{(k'_{12})^-}{\rightleftarrows}} B_2, \qquad B_2 \overset{(k'_{23})^+}{\underset{(k'_{23})^-}{\rightleftarrows}}
B_3,  \qquad B_3 \overset{(k'_{31})^+}{\underset{(k'_{31})^-}{\rightleftarrows}} B_1 \EEQ
(with arbitrary transition rates) whose graph coincides in the
zero-concentration limit
with $G_{(123)\to (1'2'3')}$ through the state identification
$(A_1,A_2,A_3,B_1,B_2,B_3)\leftrightarrow (1,2,3,1',2',3')$. 
Adding all the forward reactions $(\rightarrow)$ with
coefficients $c_{{\cal R}_1}=5, c_{{\cal R}_2}=4, c_{{\cal R}_3}=3$; $c_{{\cal R}_{1,1'}}=\half$; $c_{{\cal R}_{1'}}=\frac{5}{6}$, $c_{{\cal R}_{2'}}=\frac{4}{6}, c_{{\cal R}_{3'}}=\frac{3}{6}$  yields the result
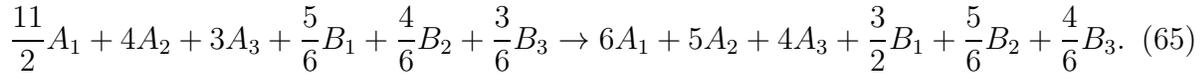
\BEQ \frac{11}{2} A_1+ 4 A_2+ 3 A_3+ \frac{5}{6} B_1+
\frac{4}{6} B_2+ \frac{3}{6} B_3 \to 6A_1+ 5 A_2+4 A_3+ \frac{3}{2} B_1+ \frac{5}{6} B_2+ \frac{4}{6} B_3.
\EEQ
Thus this network is autocatalytic in the stoechiometric sense.

\Bigskip {\bf Partial order, minimal classes, maximal classes.}  Generalizing the above example, one sees that, by
shrinking communicating classes to single points, one
reduces the oriented graph $G(M)$ to an oriented "contracted" graph
  ${\cal T}(M)$ which has no loops. (Mind that the associated {\em unoriented} graph may have loops, so that ${\cal T}(M)$ 
  is not necessarily a topological tree.) It is possible to represent this graph with edge arrows
going downwards, e.g. in the above example,

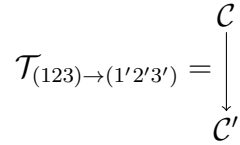
{\centerline{\begin{tikzpicture}[scale=0.75]
\draw(-2,-1) node {${\cal T}_{(123)\to (1'2'3')}=$};
\draw(0,0) node {$\cal C$};
\draw(0,-2) node {${\cal C}'$};
\draw[->](0,-0.3)--(0,-1.7);
\end{tikzpicture}}}

\Medskip with ${\cal C}=\{1,2,3\},\,  {\cal C}'=\{1',2',3'\}$. 
Note that the graph would become irreducible if (as discussed above) one added the reverse
arrow  $1'\to 1$ corresponding to the reverse reaction $2B_1\to A_1$.
We get a partial order on the set of classes by letting
${\cal C}'\succ {\cal C}$ if there is a ${\cal T}$-path downstream (i.e. following the probability flow) from
${\cal C}$ to ${\cal C}'$. Maximal (downstream) elements (here ${\cal C}'$)
are called {\em closed classes}, because they have no outgoing arrows: one cannot escape from them. Minimal 
(upstream) classes (here ${\cal C}$), on the other hand, have
no ingoing arrows.

%%%%%%%%%%%%%%

\Bigskip {\bf Autocatalysis in the stoechiometric sense.}
Let us now discuss the connection between reaction networks
and graphs. Consider a reaction network with species set ${\cal S}=\{A_1,\ldots,A_{|{\cal S}|}\}$, forward reaction set $\{1,\ldots,N\}$ and positive, mass-action reaction rates for both forward and reverse
reactions. We are particularly interested in
the limit of  small concentrations, so we distinguish:

\begin{itemize}
\item[(i)] reversible reactions $A_i\leftrightarrows A_j$ ($i\not=j$);
\item[(ii)] and  irreversible reactions $A_i\overset{k_+}{\underset{k_-}{\rightleftarrows}} s_1 A_{i'_1}+\cdots+ s_n A_{i'_n}$, with
$n\ge 1$, $s_i\in\N^*$, $\sum_{i=1}^n s_i>1$.
\end{itemize}

\Medskip Let us construct the graph  associated to the linearized time-evolution generator $M$ in the zero-concentration limit; note that the graph actually depends only on the stoechiometry matrix $\mathbb S$, not on the rates, so we can call it $G({\mathbb S})$. As discussed
in \S \ref{section:3}, in case of multiple arrow $i\to j$,
 we keep only one.

\begin{itemize}
\item[(i)] Reversible reactions $A_i\leftrightarrows A_j$ contribute to $G({\mathbb S})$ two arrows $i\to j$ and $j\to i$.
\item[(ii)]  Forward reactions $ {\cal R}: \qquad A_i\overset{k_+}{\rightarrow} s_1 A_{i'_1}+\cdots+ s_n A_{i'_n}$
($\sum_i s_i>1$) contribute to $G({\mathbb S})$
an arrow $i\to i'_{\ell}$ for each $\ell=1,\ldots,n$.
Reversible reactions $A_i\leftrightarrows A_j$ decompose
into two forward reactions $A_i\to A_j$ and $
A_j\to A_i$.

\end{itemize}

\noindent On the other hand, reverse reactions $\bar{\cal R}: s_1 A_{i'_1}+\cdots+ s_n A_{i'_n} \overset{k_-}{\rightarrow} A_i$, with
$\sum_i s_i>1$ contribute no arrow.

\Bigskip {\em Case of the "$A_1 A_2 A_3 \longrightarrow B_1 B_2 B_3$" autocatalytic kinetic reaction network.}

 The linearized evolution generator is a sum of 7 matrices,
one per reaction (provided paired generators associated to forward/reverse reversible
reactions $i\leftrightarrows j$, $i'\leftrightarrows j'$
are summed together), $M=\sum_{i} M({\cal R}_{i}) +
M({\cal R}_{11'})+
\sum_{i'} M({\cal R}_{i'})$, with 

\medskip
\begin{tikzpicture}[scale=0.8]
\draw(-2,0) node {$ M({\cal R}_{1})=$};
\draw(0,-3)--(-0.5,-3)--(-0.5,3)--(0,3);
\draw(0,0)--(6,0); \draw(3.3,3)--(3.3,-3); 
 \draw(6,-3)--(6.5,-3)--(6.5,3)--(6,3);
\begin{scope}[shift={(0.5,-0.5)}]
 \draw(0,3) node {$-k^+_{12}$}; \draw(1,3) node {$k_{12}^-$};
 \draw(0,2) node {$k^+_{12}$}; \draw(1,2) node {$-k_{12}^-$};
\draw(4,2) node {$0$}; \draw(1,-1) node {$0$}; \draw(4,-1) node {$0$};
\draw(0,4) node {$1$};  \draw(1,4) node {$2$}; 
\draw(2,4) node {$3$}; 
\draw(3.5,4) node {$1'$}; \draw(4.5,4) node {$2'$}; 
\draw(5.5,4) node {$3'$}; 
 \end{scope}  
\end{tikzpicture}

and similarly for the four other reversible generators $A_2\leftrightarrows A_3$, $B_i\leftrightarrows B_j$; 
these are probability preserving adjoint Markov generators since the sum of coefficients on any column is zero;

\medskip
\begin{tikzpicture}[scale=0.8]
\draw(-2,0) node {$ M({\cal R}_{3})=$};
\draw(0,-3)--(-0.5,-3)--(-0.5,3)--(0,3);
\draw(0,0)--(6,0); \draw(3.3,3)--(3.3,-3); 
 \draw(6,-3)--(6.5,-3)--(6.5,3)--(6,3);
\begin{scope}[shift={(0.5,-0.5)}]
 \draw(2,3) node {$2k^+_{31}$};  \draw(2,1) node {$-k^+_{31}$};
\draw(4,2) node {$0$}; \draw(1,-1) node {$0$}; \draw(4,-1) node {$0$};
\draw(0,4) node {$1$};  \draw(1,4) node {$2$}; 
\draw(2,4) node {$3$}; 
\draw(3.5,4) node {$1'$}; \draw(4.5,4) node {$2'$}; 
\draw(5.5,4) node {$3'$}; 
 \end{scope}  
\end{tikzpicture}

which is an adjoint Markov generator with negative killing
rate; and

\medskip
\begin{tikzpicture}[scale=0.8]
\draw(-3.5,0) node {$ M({\cal R}_{11'}) +
{\color{red}M(\bar{\cal R}_{11'})}=$};
\draw(0,-3)--(-0.5,-3)--(-0.5,3)--(0,3);
\draw(0,0)--(6,0); \draw(3,3)--(3,-3); 
 \draw(6,-3)--(6.5,-3)--(6.5,3)--(6,3);
\begin{scope}[shift={(0.5,-0.5)}]
 \draw(0,3) node {$ -k_{11'}^+$}; 
\draw(3.7,3) node {${\color{red} 2[B_1]k_{11'}^-}$}; \draw(0,0) node {$ 2k_{11'}^+$}; 
\draw(3.7,0) node {${\color{red}
-4[B_1] k_{11'}^-}$};
\draw(0,4) node {$1$};  \draw(1,4) node {$2$}; 
\draw(2,4) node {$3$}; 
\draw(3.5,4) node {$1'$}; \draw(4.5,4) node {$2'$}; 
\draw(5.5,4) node {$3'$}; 
 \end{scope}  
\end{tikzpicture}

for the irreversible  reaction ${\cal R}_{11'}$ coupling
${\cal C}_1$ to ${\cal C}_{1'}$, to which  one
has added the (red) reverse reaction, with rate 
proportional to the low concentration
$[B_1]$, absent in the zero-concentration limit. 

\Bigskip {\bf Path measure for generalized Markov
generators}  (see (Norris 1997, chap. 2)).  
When killing rates vanish, one has a probability law $\P$ on trajectories $(X(t))_{0\le t\le T}$: letting $t_1=0$,  $(T_k)_{k\ge 2}
\le T$ be the jumping times,
\BEA && \P[T_k=t_k+dt_k,X(T_k)=x_k, 2\le k\le \ell \ |\ X(t_1)=x_1] 
\nonumber\\
&&\qquad =
\Big[\prod_{k=1}^{\ell-1} \Big( e^{(t_{k+1}-t_{k}) M_{x_k,x_k}} 
dt_{k+1} \ \times \ 
 M_{x_{k+1},x_k} \Big)\Big] \, \times \, e^{(T-t_{\ell})M_{x_{\ell},x_{\ell}}} \label{eq:P}
\EEA
Integrating over the jumping times, one obtains the
law of the trajectories $\underline{X}=(\underline{X}_k)_{k\ge 0}$  of the underlying "skeleton" discrete-time Markov chain,
\BEQ \proba[\underline{X}_k=x_k, 2\le k\le \ell\ |\ \underline{X}_1=x_1]=
\prod_{k=1}^{\ell-1} w_{x_k\to x_{k+1}}  \label{eq:Pbar}
\EEQ
with transition rates 
\BEQ w_{i\to j}:=\frac{M_{ji}}{|M_{i,i}|}, \qquad i\not= j\EEQ
featuring the Markov generator $M^t$. 
{\em We generalize to arbitrary killing rates and use 
(\ref{eq:P}), (\ref{eq:Pbar}) as a definition for an
 unnormalized measure $\P$ over
trajectories.}

\Bigskip  {\em A path representation of the resolvent.}
When killing rates vanish, $\sum_{j\not=i} w_{i\to j}=1$, and $w_{i\to j}$ are simply the transition rates of the underlying skeleton discrete-time Markov chain; by extension, the coefficients $w_{i\to j}$  will be called {\em transition rates} in the general case. In chemical terms, $w_{i\to j}$ measures the {\em specificity} of the reaction $i\to j$.
{\em  Let  $\alpha:=$diag(($\alpha_i)_{1\le i\le |{\cal S}|})$ a positive diagonal matrix, and $M_{\alpha}:=M-\alpha$. Then
 \BEQ (R(\alpha))_{i,j}:=\int_0^{+\infty} dt\, 
(e^{tM_{\alpha}})_{i,j}\in[0,+\infty]
\EEQ
 defines a matrix
with positive coefficients, which can be computed as a sum
over backward paths $i= x_1\to x_2\to\cdots\to x_{\ell-1}\to x_{\ell}=j$ of arbitrary length $\ell\ge 0$,
\BEQ (R(\alpha))_{i,j}=\sum_{\ell\ge 0} \sum_{x_2,\ldots,x_{\ell-1}\in{\cal S}} \Big( \prod_{k=1}^{\ell-1}w(\alpha)_{x_{k+1}\to x_k}  \Big)
\, \times\, \frac{1}{|M_{j,j}|+\alpha_j}.  \label{eq:traj}
\EEQ
where
\BEQ w(\alpha)_{x_{k+1}\to x_k}:=\frac{M_{x_k,x_{k+1}}}{|M_{x_k,x_k}|+\alpha_{x_k}}.
\EEQ}

\Medskip When finite, $(R(\alpha))_{i,j}<\infty$
are the coefficients of the resolvent  $(\alpha-M)^{-1}=(-M_{\alpha})^{-1}$; see
e.g. (Revuz  1999, chap. III), or 
(Norris 1997, \S 4.2) for an introduction in connection to potential theory.

\Medskip {\em Proof.} The $\ell=0$ contribution is non-zero only if
$i=j$, in which case it corresponds to the integral
$\int_0^{+\infty} dt \, e^{t(M_{\alpha})_{j,j}} = \frac{1}{|M_{jj}|+\alpha_j}$. Splitting $M_{\alpha}$ into
$(M_{\alpha})_{diag}+ (M_{\alpha})_{off}$, where
$(M_{\alpha})_{diag}$, resp. $(M_{\alpha})_{off}=M_{off}$, 
is its diagonal part, resp. its off-diagonal (jump) part, and expanding the exponential $(e^{tM_{\alpha}})_{i,j}$ 
using the Feynman-Kac (or Trotter product) formula, one obtains a sum over trajectories 
$(x(t'))_{0\le t'\le t}$ such that
 $x|_{[t_k,t_{k+1})}=x_k$, $k=1,\ldots,\ell-1$, and $x|_{[t_{\ell},t]}=j$, with
$0=t_1<t_2<\ldots<t_{\ell}<t$.  Integrating over $t$,  one obtains  multiple integrals
\BEA && \Big(\int_0^{+\infty} dt_2\, e^{t_2(M_{\alpha})_{x_1,x_1}} M_{x_1,x_2}\Big)
\Big(\int_{t_2}^{+\infty} dt_3\, e^{(t_3-t_2) (M_{\alpha})_{x_2,x_2}}  M_{x_2,x_3}\Big)\cdots \nonumber\\
&&\qquad\qquad \Big(\int_{t_{\ell-1}}^{+\infty} dt_{\ell}\, e^{(t_{\ell}-t_{\ell-1}) (M_{\alpha})_{x_{\ell-1},x_{\ell-1}}} M_{x_{\ell-1},x_{\ell}} \Big) \  \times\ \int_{t_{\ell}}^{+\infty} dt \, e^{(t-t_{\ell})(M_{\alpha})_{x_{\ell},x_{\ell}}}, \nonumber\\
\EEA
yielding (\ref{eq:traj}).

\Medskip \begin{Lemma}[Properties of the resolvent for $M$ irreducible] \label{lem:resolvent}
  {\em We  assume that $M$ is irreducible.} Then:

\begin{itemize}
\item[(i)] {\em Coefficient functions
$\alpha\mapsto (R(\alpha))_{i,j}$ are decreasing,} namely, if $\alpha\le \alpha'$, i.e.
$\alpha_i\le \alpha'_i$ for all $i$, then $R(\alpha')\le R(\alpha)$;

\item[(ii)] let $t\mapsto \alpha(t)$ $(t\ge 0)$ be an increasing
function of time, i.e. $(t'\le t)\Rightarrow(\alpha(t')\le \alpha(t))$; then
there exists a {\em transition time} $t=t_0$  (possibly, $t_0=0$) such that all coefficients
of $R(\alpha(t))$ are $<\infty$ if $t> t_0$, and
 all coefficients
of $R(\alpha(t))$ are $\infty$ if $0\le t< t_0$.  If $t>t_0$, then
$R(\alpha(t))=(-M_{\alpha(t)})^{-1}$. 

\item[(iii)]  {\em (Lyapunov exponent)} Let $\lambda_{max}:= 
\max\{\Re(\lambda) \ |\ \lambda$ eigenvalue of $M\}$ be the
Lyapunov exponent of $M$. If one lets $\alpha(t):=t\Id$,
and the transition time $t_0$ is $>0$, 
then $\lambda_{\max}=t_0$. Conversely, if $t_0=0$, then
$\lambda_{max}\le 0$. 

\item[(iv)] {\em (positivity criterion for Lyapunov exponent)}
assume $R(\alpha)=+\infty$ for some $\alpha\ge 0$ which is
not identically zero, then $\lambda_{max}>0$.

\end{itemize}
\end{Lemma}

\Medskip {\em Proof.} For (ii) we need only remark that
$\Big(\exists i,j,\, (R(\alpha))_{i,j}=+\infty\Big)\Rightarrow
\Big(\forall i,j,\, (R(\alpha))_{i,j}=+\infty\Big)$. Namely, let
$i',j'$ be indices;
$M$ being irreducible, there exists a backward path from $i'$ to $i$, and a backward path from $j$ to $j'$; 
sandwiching $(R(\alpha))_{i,j}$ -- which is the sum of the
weights of all backward paths from $i$ to $j$ -- between
them, one gets $(R(\alpha))_{i',j'}=+\infty$. 

\Medskip Let now $\alpha(t)=t\Id$. If $t>\max(0,\lambda_{max})$, then (as can be proved by standard arguments using e.g. Jordan's form for $M$) there
exists some constant
$c>0$ s.t. for all $\tau>0$, 
$ ||| e^{\tau M_{\alpha(t)}} |||=O(e^{-c\tau})$ 
$(|||\ \cdot\ |||$ being any norm), hence 
$R(\alpha(t))_{ij}<\infty$ for all $i,j$. Conversely, 
if  $R(\alpha(t))_{ij}<\infty$ for all $i,j$, then 
$\lambda-M$ is invertible if $\Re\lambda\ge \alpha(t)$, as 
follows from the path representation (\ref{eq:traj}). This implies
(iii).

\Medskip Discussing finally
 (iv), assume that $R(\alpha)=+\infty$ with $\alpha_i>0$.
 Let ${\cal W}_i(\alpha)$ be the  weight of {\em excursions}
 from $i$, i.e. the total weight of all backward paths
 $i=x_1\to x_2\to \cdots\to x_{\ell-1}\to x_{\ell}=i$ such
 that $x_2,\ldots,x_{\ell-1}\not=i$. Then
 
\BEQ (R(\alpha))_{i,i}=\Big( \sum_{n=0}^{+\infty} ({\cal W}_i(\alpha))^n \Big) \ \times\ \frac{1}{|M_{i,i}|+\alpha_i}
\EEQ

hence
\BEQ (R(\alpha)=+\infty)
 \Rightarrow ({\cal W}_i(\alpha)\ge 1).
\EEQ
 The function
 $\alpha\mapsto {\cal W}_i(\alpha)$ is a strictly decreasing function,
 so ${\cal W}_i(\alpha/2)>1$. This strict inequality remains true (by continuity)
 in a neighborhood $\Omega$ of $\alpha/2$ in $\R_+^{{\cal S}}$, and (by monotony) for
 all $\alpha'$ such that $\alpha'\le\alpha''$ for some
 $\alpha''\in\Omega$. In particular, for $t$  small enough,
 $R(t\Id)=+\infty$, so that $\lambda_{max}>0$.  \hfill \eop

\Medskip To finish with, we study in some details
the onset of the exponential growth using the Perron-Frobenius
theorem. 

\begin{Lemma} \label{lem:PF}
 
\begin{itemize}
\item[(i)] If $\lambda$ is an eigenvalue of $M$ and
$\Re\lambda=\lambda_{max}$, then $\lambda=\lambda_{max}$. In particular, $\lambda_{max}$ is an eigenvalue of $M$. 
Furthermore, the multiplicity of $\lambda_{max}$ is $1$, 
and there exists an associated eigenvector with $>0$ coefficients. 
\item[(ii)] Let $\tau>0$.  There exist two constants $c=c(\tau),C=C(\tau)>0$ such that,
for every nonzero initial concentration vector  $v$ with $\ge 0$ coefficients, and for every $t>\tau$,
\BEQ c\Big(\max_i v_i\Big)\,  e^{\lambda_{max}t} 
 \le   \Big(e^{tM} v\Big)_i \le C\Big(\max_i v_i\Big) \, e^{\lambda_{max}t}.  \label{eq:lem6.2.2}
 \EEQ 
\end{itemize}

\end{Lemma}

\noindent The upper bound (\ref{eq:lem6.2.2}) holds uniformly in $\tau$, but the lower bound degenerates as $\tau\to 0$
(because $(e^{tM} v)_i\to_{t\to 0} v_i$ can vanish). The
{\em homogenization time} $\tau$ is discussed below.

\Medskip {\em Proof.} 
\begin{itemize}
\item[(i)] This is  a consequence
of the Perron-Frobenius theorem, since (for $C>0$ large
enough constant) $M+C\Id$ has
positive coefficients and is irreducible.

\item[(ii)] Fix the eigenvector $v_{max}$ associated to the 
maximal eigenvalue $\lambda_{max}$ by requiring that
$v_{max,i}>0$ for all $i$ and $||v_{max}||_{\infty}:=\max_i v_{max,i} =1$.  The upper bound follows by standard
computations from splitting $v$ into $v_{//}+w$, where
$v_{//}$ is the linear projection of $v$ onto the
one-dimensional eigenspace $\R v_{max}$ parallel to
the sum of all other generalized eigenspaces.  For the lower 
bound, we note that $(e^{\tau M})_{ij}>0$ for all indices $i,j$ and
$\tau>0$. Fix some (small) instant $\tau>0$; there exists
then $c>0$ such that that 
\BEQ (e^{\tau M} v)_i \ge c||v||_{\infty} v_{max,i}.
\label{eq:homoeneization-time}
\EEQ
 Let $t\ge\tau$. Since the matrix $e^{(t-\tau)M}$ has positive coefficient, we get
$(e^{tM}v)_i \ge c ||v||_{\infty} \, \Big( e^{(t-\tau)M}v_{max}\Big)_i = c||v||_{\infty}\, e^{(t-\tau) \lambda_{max}}
v_{max,i}.$
\end{itemize} \hfill \eop

\Medskip For applications, we are mostly interested
in the onset of the exponential growth regime, and may
assume that $\lambda_{max}>0$.  Let $M:=M([A]=0)$ be the generalized Markov generator obtained by linearizing the kinetic equations at zero
concentrations.  By definition, $\frac{d[A]}{dt}=M[A]+O([A]^2)$.  Hence it follows from the above Corollary
that, for all $i\in {\cal S}$ and $t>\tau$,  
\BEQ \frac{[A_i](t)}{\max_j \, ([A_j](t=0))}  \approx  e^{\lambda_{max}t}   \label{eq:spontaneous}
\EEQ
for   time values $t$ such that $\max_i \, [A_i](t=0) 
\, \times \, e^{\lambda_{max}t} $ is small enough (depending
on kinetic rates), where $a\approx b$ $(a,b>0)$ means: $ca<b<Ca$ for constants $c,C>0$  independent of $t$, $i$  and $[A](t=0)$, but depending on the homoegeneization time $\tau$.  Eq. (\ref{eq:spontaneous}) may
be regarded as a mathematical expression for {\em spontaneous autocatalysis}. The homoegeneization time $\tau$ should
be chosen as small as possible in order for 
(\ref{eq:homoeneization-time}) to hold for a not too
small constant $c$,  with $M=M([A]=0)$.

%%%%%%%%%%%%%%%%%%%%%%%%%%%%%%%%%%%%%

%%%%%%%%%%%%%%%%%%%%%%%%%%%%%%%

\end{document}